\documentstyle[aps,epsf,array]{revtex}
\def\npb#1{ {\sl Nucl. Phys.} {#1}}
\def\gsim{\mathrel{\raise.3ex\hbox{$>$\kern-.75em\lower1ex\hbox{$\sim$}}}}
\def\lsim{\mathrel{\raise.3ex\hbox{$<$\kern-.75em\lower1ex\hbox{$\sim$}}}}
\def\arrl{\mathrel{\raise.3ex\hbox{$\longrightarrow$\kern-1.7em\lower1ex
       \hbox{$\scriptstyle t\rightarrow\infty$}}}}
\def\arra{\mathrel{\raise.3ex\hbox{$\longrightarrow$\kern-1.7em\lower1ex
       \hbox{$\scriptstyle \alpha\rightarrow0$}}}}
\def\partder#1#2{\frac{\partial #1}{\partial #2}}
\def\llangle{\langle\langle}
\def\rrangle{\rangle\rangle}

\def\jsp{{\sl  J. Stat. Phys.}}

\def\dof{\varphi}
\draft
\begin{document}

\vspace*{1cm}

\begin{center}
{\Large {\sc  Non-Equilibrium Statistical Mechanics of\\
  Classical Lattice $\phi^4$ Field Theory}}

\vspace{1.3cm}

Kenichiro Aoki\footnote{ E--mail: {\tt ken@phys-h.keio.ac.jp}}
{\rm and} Dimitri Kusnezov\footnote{E--mail: 
  {\tt dimitri@nst4.physics.yale.edu}} \\
$^*${\small\sl Dept. of Physics, Keio University, 
  4---1---1 Hiyoshi, Kouhoku--ku, Yokohama 223--8521, 
  Japan}\\
$^\dagger${\small\sl Center for Theoretical Physics, Sloane Physics
  Lab, Yale University, New Haven, CT\ 06520-8120}
\vskip 1.8 cm \parbox{14.5cm} {
  \begin{center}
    \large\sc ABSTRACT
  \end{center} 
  {\hspace*{0.3cm} Classical $\phi^4$ theory in weak and strong
    thermal gradients is studied on the lattice in (1+1) dimensions.
    The steady state physics of the theory is investigated from first
    principles and classified into dynamical regimes.  
    We derive the bulk
    properties associated with thermal transport, and explore in
    detail the non-equilibrium statistical mechanics of the theory as
    well as connections to equilibrium and irreversible
    thermodynamics. 
    Linear response predictions are found to be valid for systems
    quite far from equilibrium and are seen to eventually break down
    simultaneously with local equilibrium.
    }}
\end{center}
\vspace{5mm}

\section{Introduction}
\label{sec:intro}
The physics of non-equilibrium systems includes a broad class of
phenomena, such as the physics of steady states, relaxation and
dynamics far from equilibrium. Dynamical processes which range from
those in the early universe, to ultra-relativistic heavy ion
collisions and the formation of the quark-gluon plasma all involve
non-equilibrium physics in an essential manner, perhaps requiring an
understanding of the physics beyond linear response.  Physics of
steady states contains interesting non-equilibrium phenomena, such as
transport, spatially varying observables, and is also important to the
study of systems where the time scales for local equilibration are
smaller than macroscopic time scales which might describe some global
evolution of the system.  Non-equilibrium steady states can be
realized in many ways, such as the placement of a systems in a thermal
gradient, or in an environment which provides shearing, pressure
gradients and so forth.  In this article, we will examine the
statistical mechanics of the non-equilibrium steady states of a
classical field theory in thermal gradients.  This will allow us to
understand the behavior of the theory under these non-equilibrium
conditions and to consider problems related to the range or validity
of local equilibrium, linear response, equilibrium thermodynamics and
statistical mechanics.  In particular, the thermal gradients we study
can be quite strong and in such situations, it is natural to ask
whether the system always relaxes to local equilibrium.

When systems are near equilibrium, one expects linear response to
provide a description of the transport coefficients. However, there is
no means to address its regime of validity within the theory itself.
Furthermore, even in this regime, in low dimensions, $d\leq2$, it has
been argued using kinetic theory that linear response often does {\it
  not} hold\cite{tail}; the Green-Kubo autocorrelation functions are
expected to behave as $t^{-d/2}$, leading to divergent transport
coefficients.  
Such divergences of the Green-Kubo integral have been observed in
certain low dimensional systems such as the FPU model \cite{lepri} or
the diatomic Toda lattice \cite{hatano}, and seem to be endemic to
Hamiltonians which conserve total momentum.  Other studies have also
found thermal transport in the linear regime to diverge
\cite{fpu-later,mareschal}\ or have focused on somewhat exotic models
\cite{casati,finite-1d-transport}, while strong thermal gradients have
been studied in cellular automata\cite{takesue}.

In this work, we attempt to address many of the basic questions
regarding the non-equilibrium properties of the $\phi^4$ theory on the
lattice in (1+1) dimensions, by studying the model both near and far
from equilibrium.  We choose the $\phi^4$ theory since it is a
prototypical model which appears in a variety of contexts, including
particle physics.  From the outset, we should point out that our work
has two limitations, namely that it is classical and that it is a
lattice field theory.  On the other hand, we make no further
approximations and we analyze the model from first principles without
any dynamical assumptions.  This will allow us to answer interesting
physical questions that cannot yet be addressed in the full quantum
case.  Furthermore, the approach we adopt can be generalized to other
classical lattice field theories in a straightforward manner.  Our
main objective is to develop a comprehensive understanding of the
underlying dynamics of the scalar field theory in thermal gradients
and to lay the ground work for further analysis.  As such, we shall
provide the necessary details of our methods for further work and in
such a manner so that can be easily generalized to other models.

Classical field theory is relevant to high temperature behavior of
quantum field theories: For instance, recently, it was used to derive
properties of the standard electroweak model at finite temperature
\cite{EW,AS}.  Properties of the quantum $\phi^4$ theory in
equilibrium has also been studied previously \cite{hosoya,yaffe}.
However, the relation between the physical quantities in the classical
lattice field theory and the quantum theory is far from trivial and it
is beyond the scope of this work.  The dynamics of classical field
theories is of interest on its own right, which in our case can be
thought of as the dynamics of an anharmonic chain.
While we discuss the results of the (1+1) dimensional theory in this
work, it is {\it not an essential limitation} of our model or
approach; we do find that the basic physics understanding developed in
(1+1) dimensions carries over to results in (3+1) dimensions, although
the latter will be discussed elsewhere.  As such, our analysis in
(1+1) dimensions is not specific to 1-d systems.  It should be noted
that classical $\phi^4$ theory has been considered in various contexts
in the past: The Lyapunov spectra has been computed in the
microcanonical ensemble in massless\cite{gong} and
massive\cite{caiani} models.   Equilibration of the model was studied
in \cite{parisi}. 

In addition to elucidating the physics underlying some of the
non--equilibrium phenomena of field theories, we believe that our
results shed light on the nature of non-equilibrium statistical
ensembles. The extension of the Gibbs ensemble to the non-equilibrium
steady state remains unanswered. However, limited results to date
suggest that the theory is far from trivial,  including divergent Gibbs
entropy and singular steady-state measures\cite{revs}.  While some
approaches exist, such as maximum entropy or projection techniques to
construct non-equilibrium operators\cite{zubarev}, assumptions must be
made about the non-equilibrium state in order to compute its
properties.
By using classical field theory as our starting point, we can use
existing techniques to construct non-equilibrium steady states without
assumptions on the dynamics of the model which are symptomatic to
other approaches.  A seemingly simple question concerns the thermal
profile $T(x)$ which develops in a system when it is in a thermal
gradient. In other approaches, often some form is assumed for the
profile $T(x)$, which in principle should be dynamically obtained, as
we shall do here.  Furthermore, in our study, whether local
equilibrium is achieved is {\it not} an assumption but is determined
{\it dynamically} by the system.  We will see that there are
qualitative differences between the behavior of the system in the
various regimes. These are characterized in Table 1.

We would like to emphasize once again the motivation for this
analysis: While the lattice model we work with does have well defined
thermal transport, a single component scalar {\it continuum} field
theory does not support thermal conductivity in the usual sense.  
It would be interesting to generalize to more complex theories with more 
than one
conserved current, such as the two component scalar field
theory, gauge theory with matter and so on.  On the other hand, rather
than introduce additional degrees of freedom and observables such as
charge or matter density, we prefer first to focus on some important
questions in the lattice theory.  By focusing on the simplest lattice
theory, we are able to clearly present an approach to the
non-equilibrium statistical mechanics of classical lattice models from
first principles, which can then be applied to other models like the
ones mentioned above.  Our results are also  of interest to the
statistical mechanics of many body systems in non-equilibrium.
So as is, we do not attempt to make claims concerning the continuum
limit  but rather concentrate on elucidating the physics of
the lattice theory near and far from equilibrium.

In section 2, we describe the model we study and how we analyze the
theory, particularly paying attention to the way the temperature
boundary conditions are implemented.  In section 3, we discuss thermal
transport in our theory.  In section 4, we analyze the equilibrium
physics of the model and in section 5, the non--equilibrium physics.
In particular, we study the thermal conductivity and its temperature
dependence.  We analyze the thermal profiles and establish that the
linear response theory works even for visibly curved profiles, up to
certain thermal gradients.  We further examine the relations between
the various physical quantities such as entropy, speed of sound, heat
capacity and the thermal conductivity.  We end with a discussion in
section 6.

\begin{center}
  Table 1: Behavior of the $\phi^4$ theory under varying thermal
  gradients. \\
  Here, $\ell$ is the mean free path as explained in section
  \ref{sec:non-eq}.  \\  
  \vspace{0.3cm} \leavevmode {\sl
    \begin{tabular}{lcl}
      \hline
      Regime      &$\qquad$  &  Properties\\
      \hline
      Global Equilibrium (GE)       
      &&  $\nabla T = 0$, $f(\pi,\phi)\sim\exp[-H/T]$.\\
      Local Equilibrium I (LE-I) 
      &&       $\Delta T/T\ll1$;  
      $\nabla T =$constant; \\
      &&Fourier's law holds globally;\\
      && Agreement with linear response theory.\\
      Local Equilibrium I$\!$I (LE-I$\!$I) && 
      $\ell\nabla T/T\alt 1/10$; $ \nabla T \not=$constant;\\ & &
      Fourier's law holds locally;\\ && Small deviations from linear
      response theory;\\ && Existence of boundary temperature jumps.
      \\ 
      Local Non-Equilibrium (LNE) 
      &&   $\ell\nabla T/T\agt 1/10$; 
      $ \nabla T \not=$constant;\\
      && Local equilibrium description inadequate;\\
      &&  Definition of temperature  ambiguous;\\
      && No suitable definition for transport coefficients.\\
      \hline
    \end{tabular}}
\end{center}

\section{The Model}
We start with the $\phi^4$ Lagrangian (with the metric convention
$(-,+)$),
\begin{equation}  \label{continuum-lagrangian}
  -{\cal L}= {\frac{1}{2}} \left(\frac{\partial\tilde\phi
      (\tilde x)} {\partial\tilde x_\mu} \right)^2
  +{1\over2}\tilde m^2\tilde\phi(\tilde x)^2
  +{\frac{\tilde g^2}{4}} \tilde\phi(\tilde x)^4 .
\end{equation}
We discretize and perform the rescaling 
\begin{equation}
  \phi_{x}(t)=a\tilde g\tilde \phi(\tilde{x},\tilde t ), \quad
  t=\tilde t/a, \quad {x}=\tilde {x}/a, \quad
  m^2=\tilde m^2a^2,
\end{equation}
where $a$ is the lattice spacing.  We then obtain the
corresponding Hamiltonian where the lattice spacing is scaled
out 
\begin{equation}  \label{lattice-hamiltonian}
  H(\pi,\phi)=\frac{1}{2}\sum_{i}\left[\pi_{i}^2 +
    \left(\nabla \phi_{i}\right)^2 +  m^2 \phi_{i}^2 
       + \frac{1}{2}\phi_{i}^4\right].
\end{equation}
Here ${k}=1,2,\ldots L$ runs over all sites in the lattice, 
the lattice derivative is $\nabla\phi_{k}\equiv \phi_{k +
  1}-\phi_{k}$.
The resulting equations of motion are:
\begin{equation} \label{microcan}
 \dot\phi_i = \pi_i,\qquad \dot\pi_i =  (\nabla^2\phi)_i
 - m^2\phi_i - \phi_i^3.
\end{equation}
Here, we defined the lattice Laplacian as
$(\nabla^2\phi)_k\equiv \phi_{k+1}-2\phi_{k}+\phi_{k-1}$.

\subsection{ Finite Temperature Equilibrium Dynamics: $\nabla T=0$}

Starting from the microcanonical dynamics (\ref{microcan}), we can
develop a realization of the constant temperature dynamics using the
global demons of Ref. \cite{thermostats}. In this approach, auxiliary
variables are added to the systems which dynamically emulate the
presence of a heat bath.  This type of dynamics, while not as
rigorously understood, has been shown to converge much faster than the
optimized hybrid Monte Carlo methods.  Further, it does not suffer as
much from critical slowing down\cite{js}.
   
When we are interested in studying the statistical properties of
a system described by an action $S(\dof)$, where
$\dof=(\dof_1,\dof_2,\ldots\dof_n)$ are the degrees of freedom,
we usually start with the definition of a statistical measure,
such as
\begin{equation}
  f {\cal D}\mu(\dof) \sim \exp[- S(\dof)/T] {\cal
    D}\mu(\dof).
\end{equation}
Here, ${\cal D}\mu(\dof)$ might include constraints in the dynamical
space, as in the case of motion on curved manifolds, such as Lie
groups. Because we know the measure, steady-state values of
observables are readily determined. For an arbitrary observable ${\cal
  O}$, we have
\begin{equation}
  \label{eq:ens}
  \langle{\cal O}\rangle = \frac{1}{ Z}\int\;{\cal
    D}\mu(\dof)\;e^{- S(\dof)/T}{\cal O},\qquad Z = \int\; {\cal
    D}\mu(\dof)\; e^{- S(\dof)/T}.
\end{equation}
While the approach we discuss is suited to general measures, in
this article we use the measure ${\cal D}\mu(\dof)={\cal D}\dof$
over the phase space, where $\dof$ will typically represent
canonically conjugate coordinates and momenta,
$\dof=(\phi,\pi)$, and $S(\dof)$ is taken as a Hamiltonian,
$S(\phi,\pi)=H(\phi,\pi)$.
While the dynamics of the model may now be easily implemented using
equations of motion, (\ref{microcan}) --- often referred to as the
molecular dynamics method --- we would like to add finite temperature
{\sl constraints} to the equations. For this to happen, we must no
longer evolve on the constant energy surface, so that $S(\dof)$ should
no longer be conserved. The method we discuss here is reminiscent of
Parisi and Wu's stochastic quantization\cite{parisi-wu}, although the
one adopted in our work is deterministic and time-reversal invariant.
It is also a versatile approach in that it has been applied to systems
with non-trivial measures, such as Lie algebraic Hamiltonians,
equilibrium and non-equilibrium quantum systems, atomic clusters and
molecules, magnetic materials and lattice models\cite{apps}.

There are many formulations of this dynamics, initially
motivated by the approaches of Nos\'e and Hoover\cite{nose-hoover}.
Consider the following equations of motion for a thermostatted
site labeled by $k$:
\begin{equation}
 \dot\phi_k = \pi_k ,\qquad
 \dot\pi_k = -\partder{S}{\phi_k} 
       - \frac{dG(w_k)}{dw_k}\; F(\pi_k)     
       - \frac{dG'(w'_k)}{dw'_k}\; F'(\pi_k).
\end{equation}
We have added two additional degrees of freedom, $w_k,w'_k$, which
couple through forces indicated above.  These extra degrees of
freedom, so called `demons', may be coupled either to the fields,
$\phi_k$, or to their `momenta', $\pi_k$.  Here, we choose to couple
them only to $\pi_k$'s in order to have the ability to apply
thermostats locally at any one site.  The microcanonical limit is
recovered when these extra degrees of freedom are decoupled.
In this extended space, $\dof=(\phi_i,\pi_i,w_k,w'_k)$ , we define
a new action which is the old one plus additional terms for the
demons:
\begin{equation}
  \label{eq:meas}
  f(\phi,\pi,w,w') = \exp\left( -\left[
      S(\phi,\pi) + \sum_{k:\ \rm thermostatted\atop  sites}
      \left(G(w_k)+G'(w'_k)\right)\right]/T\right) .
\end{equation}
In contrast to microcanonical dynamics, where the Hamiltonian is a
constant of the motion, $f$ is not preserved by the constant
temperature dynamics. While the choice of forces in the equations of
motion as well as that of $f$ are seemingly arbitrary, steady state
expectation values will be independent of these under reasonably
general conditions. Consequently, they are chosen to optimize
convergence of the physical variables.

To find the dynamics associated with the demons $w_i,w'_i$, we simply
require that $f$ satisfy a continuity (Liouville) equation in the
configuration space $\dof=(\phi_i,\pi_i,w_k,w'_k)$:
\begin{equation}
  \label{eq:prob}
  0 = \partder{f}{t} + \sum_{i}
  \partder{(\dot{\dof}_i f)}{\dof_i}.
\end{equation}
This is equivalent to requiring that the master equation,
enforcing conservation of probability under evolution of the
ensemble, be satisfied.  By substituting the equations of motion
into the continuity equation, and using the definition of
$f$, we can derive solutions  for $\dot w_k,\dot w'_k$:
\begin{equation}
  \label{eq:grd}
  \dot{w_k} = \pi_k F(\pi_{k})-T{d{F(\pi_{k})}\over{d\pi_k}}
             ,\qquad
  \dot{w'}_k = \pi_k F'(\pi_{k})-T{d{F'(\pi_{k})}\over{d\pi_k}}.
\end{equation}
By construction, this dynamics preserves the measure
Eq.~(\ref{eq:meas}), so that time averages of observables on a given
trajectory will converge to the configuration space average over the
canonical measure.  There is clearly some freedom in defining the
dynamics; namely, the functions $G(w),G'(w')$ and $F(\pi),F'(\pi)$.
The only restriction on $G(w),G'(w')$ is that the measure
Eq.~(\ref{eq:meas}) leads to a finite integral; in general the
auxiliary variables $w$ can have any desired measure. In practice,
highly non-linear functions are impractical since they will require
small integration time steps.  For these reasons, it is convenient to
take $G(w),G'(w)$ to be $\mu w^2/2$ or $\mu'w^4/4$, where $\mu,\mu'$
are positive constants.  The constant couplings $\mu,\mu'$ of the
demons to the physical degree of freedom are in principle arbitrary as
long as the phase space integration is finite.  
Choosing these couplings to be too weak will make the convergence slow
or choosing them to be too strong will lead to small time steps in the
evolution, so that the couplings are chosen to optimize the
convergence of physical observables.  A necessary condition for
$F(\pi),F'(\pi)$, on the other hand, is for it to be at least linear
in its argument, the minimal requirement for the existence of the
fluctuations in the phase space volume which allows for the
exploration of the canonical measure.
The precise relation to the fluctuations in a phase space volume
${\cal V}$, or equivalently, the instantaneous phase space
compressibility, can be found using the divergence theorem
\begin{equation}
  \label{eq:volf}
     {1\over{\cal V}}
     \frac{d{\cal V}}{ dt} =-\int_{\cal V} {\cal D}\varphi\;
     \sum_{k}\left\langle\frac{dG(w_k)}{ dw_k}
       {d{F(\pi_k)}\over d{\pi_k}} + 
       \frac{dG'(w'_k)}{dw'_k}{d{F'(\pi_{k})}\over d{\pi_k}}
     \right\rangle,
\end{equation}
where the index $k$ runs over the thermostatted sites.
In this paper, we do not explore the effect of different choices
of $G(w),G'(w),F(\pi),F'(\pi)$.  Such studies have been done on
various other systems \cite{apps}.
 
Finally, we note that the linearized equations of motion are
evolved by the stability matrix, $\partial \dot \dof_i/\partial
\dof_j$. The eigenvalues of this matrix are the Lyapunov
exponents for the system. Hence we have the relation that
\begin{equation}
   \sum_i\lambda_i =
   \sum_i\left\langle\partder{\dot{\dof}_i}{\dof_i}\right\rangle.
\end{equation}
For the canonical ensemble, the Liouville equation gives $\sum
\lambda_i=0$, while in steady state non-equilibrium systems,
$\sum \lambda_i<0$.

One specific realization of the finite temperature equilibrium
dynamics we use couples the thermostats only at the endpoints of
the systems, $k=1,L$.  With the choice of $G(w)=w^4/4,
G'(w')=w'^2/2, F(\pi)=\pi/T,F'(\pi)=\pi^3/T$, we obtain
\begin{equation}
 \label{eq:eom}
 \begin{array}{lcl}
   \dot\phi_k &=&\pi_k \qquad k=1,2,\ldots L\\
   \dot\pi_k &=& \cases{
   (\nabla^2\phi)_k - m^2\phi_k - \phi_k^3  & $k=2,3,\ldots,L-1$\cr
   (\nabla^2\phi)_k - m^2\phi_k - \phi_k^3 
   - w_k^3\pi_k/T - w'_k\pi_k^3/T &
   $k=1,L$\cr   }\\
   \dot w_k &=& \pi_k^2/T-1,\qquad \dot w'_k =
   \pi_k^4/T-3\pi_k^2\qquad k=1,L.
 \end{array}
\end{equation}
Just thermostatting the two boundary points is sufficient to
thermalize the entire system. We then can examine the interior system
far from the boundaries to study the finite temperature theory.
Either free or fixed boundary conditions were used for the $\phi$
field with no significant effect on the physics behavior of the
theory.  To compute observables, we use the fact that in this type of
dynamics, the time averages converge to the ensemble average using the
desired ensemble (\ref{eq:meas}):
\begin{equation}
  \overline{{\cal O}} =
  \lim_{t\rightarrow\infty}\frac{1}{t}\int_0^t\!\!dt\,
  {\cal O}(\phi(t),\pi(t)) = \langle {\cal O}\rangle_{EQ} =
  \frac{\int {\cal D}\dof {\cal O} f}{\int  {\cal D}\dof f}
  \quad.
\end{equation}
\subsection{Non-equilibrium Boundary Conditions: $\nabla
  T\not=0$} 
One of the problems immediately encountered in the study of
non-equilibrium systems is the nature of the steady state statistical
distribution. While equilibrium statistical mechanics is well
understood, once we apply thermal gradients to the system, much less
is known about the system.  To set up the non-equilibrium molecular
dynamics, we use the demons to thermostat the end-points of our system
in the same way we did in the equilibrium simulation. The only
difference is that we now control the two endpoint temperatures
separately.  A consequence will be that the phase space distribution,
$f(\phi,\pi,t)$, will evolve to a non-smooth function that describes
the non-equilibrium steady state.

We take the equations of motion at finite $T$ and now introduce two
temperatures $T_1^0$ and $T_2^0$. The superscript is needed to
distinguish the thermostatted temperatures from those measured just
inside the system, which can suffer from boundary jumps which we will
analyze in detail\cite{pla}. One set of equations, similar to the equilibrium
case in Eq.~ (\ref{eq:eom}), we have used are
\begin{equation}\label{noneqeom}
 \begin{array}{lcll}
   \dot\phi_k &=&\pi_k \qquad k=1,2,\ldots L\\
   \dot\pi_k &=& \cases{
   (\nabla^2\phi)_k - m^2\phi_k - \phi_k^3  & $k=2,3,\ldots,L-1$\cr
   (\nabla^2\phi)_1 - m^2\phi_1 - \phi_1^3 
   - w_1^3\pi_1/T_1^0 - w'_1\pi_1^3/T_1^0 &   \cr   
   (\nabla^2\phi)_L - m^2\phi_L - \phi_L^3 
   - w_L^3\pi_L/T_2^0 - w'_L\pi_L^3/T_2^0 &   \cr   
   }\\
   \dot w_1 &=& \pi_1^2/T_1^0-1,\qquad \dot w'_1 =
   \pi_1^4/T_1^0-3\pi_1^2&\\
   \dot w_L &=& \pi_L^2/T_2^0-1,\qquad \dot w'_L =
   \pi_L^4/T_2^0-3\pi_L^2.&
 \end{array}
\end{equation}
One can see that the thermostats are applied to the endpoints $k=1$
and $k=L$. It should be noted that inside the boundaries, the dynamics
of the system is that of the $\phi^4$ theory itself with {\it no}
other degrees of freedom.  We have also considered both variations of
these thermostats, such as different forms of the interactions or
increasing the number of sites at each end where we apply the demons.
We will comment when these distinctions are relevant.  

The equations of motion are solved on a spatial grid, using two
methods: fifth and sixth order Runge-Kutta, and leap-frog
algorithms\cite{numrec}. We used from $10^6$ to $10^9$ time steps of
$dt$ from $0.1$ to $0.001$, with observables being sampled every
$\Delta t=20\sim100\,dt$.  The lattice size was varied from $L=20$ to
8000.  

A consequence of the non-equilibrium steady state is that the measure
becomes singular with respect to the Liouville measure \cite{revs}. We
do not rely explicitly on the singular nature of the non-equilibrium
measure.  Rather, we use it to interpret certain observables in the
non-equilibrium state.  To see this we start with the two main
equations for $f(\phi,\pi,t)$; the continuity equation and the
expression for the total derivative of a phase space valued function.
If we denote the vector $\dof$ to include all degrees of freedom,
$\dof=(\phi,\pi,w)$, we have
\begin{equation}\label{cont}
  \frac{df}{dt} =
  \frac{\partial f}{\partial t} + \sum_i
  \dot \dof_i \frac{\partial f} {\partial \dof_i}
  \quad.
\end{equation}
By combining this equation with the continuity equation
(\ref{eq:prob}), which holds both in equilibrium and in
non-equilibrium, we derive
\begin{equation}
  \frac{df}{dt} =  -f  
  \sum_i \frac{\partial \dot \dof_i}{ \partial \dof_i}.
\end{equation}
Solving for $f$, we obtain
\begin{equation}
  f(t) = f(0)\exp\left(- \int_0^t\!\!dt\,
    \sum_i \frac{\partial \dot \dof_i}{ \partial
                        \dof_i}\right)
     = f(0) \exp\left( -t \sum_i\left\langle   
         \frac{\partial \dot \dof_i}
         {\partial \dof_i}\right\rangle_{NE}\right)
     = f(0) \exp\left( -t\sum_i\lambda_i\right).
\end{equation}
In these steps we have replaced the time average with the
non-equilibrium ensemble average. The average of the divergence of the 
equations of motion is nothing more than the sum of Lyapunov
exponents. In non-equilibrium steady states, whether from thermal
gradients, or shearing, and so forth, the sum of the exponents is
observed to become negative, signaling the presence of a fractal
dimension. In the steady state,
\begin{equation}
  f(t) \arrl \infty.
\end{equation}
As the continuity equation is satisfied, the allowed phase space
volume must be shrinking onto a set of measure zero, with respect to
the original measure.
A consequence of the divergence of the distribution function in the
non-equilibrium steady state is that Gibbs entropy will also diverge,
\begin{equation}
  S_G = -\langle \log f\rangle \rightarrow -\infty.
\end{equation}
Although we do not know how to properly define the fractal
measure for our non-equilibrium steady state, we are able to
compute non-equilibrium expectation values with respect to it
using time averages:
\begin{equation}
  \langle {\cal O}\rangle_{NE}=\overline{{\cal O}}
  = \lim_{t\rightarrow\infty}\frac{1}{t}
  \int_0^t\!\!dt\, {\cal  O}(\phi(t),\pi(t)).
\end{equation}
We will verify this in the linear response regime where
near-equilibrium results can be compared to thermal equilibrium
predictions obtained using the linear response theory.

Simulations of many-body systems in non-equilibrium steady states have
found that the steady state measure is typically singular with respect
to the original equilibrium measure. As one moves further from thermal
equilibrium, the available phase space contracts onto an ergodic
fractal: the accessible points are dense in the phase space, but
fractal in nature. The resulting loss of dimension is related to the
transport coefficient. One can see that in our dynamics as well since
$f$ satisfies the continuity equation, so that total probability is
conserved, yet Eq. (19) is satisfied. This means that the accessible
phase space volume is contracted onto a set of measure zero. This type
of `dimensional loss' in the steady state can be more rigorously seen
in low dimensional systems like the Lorentz gas.

{}From the point of view of dynamical systems theory, it is possible to
understand the steady state measures under certain special conditions,
namely that the dynamics is hyperbolic or Anosov. Unfortunately the
conditions for hyperbolicity are not satisfied for our system since
the number of positive Lyapunov exponents can vary along a trajectory.
Nevertheless, it is useful to note that for hyperbolic systems
described by some flow, $x(t)=S_t x$, for time evolution operator
$S_t$ and initial point $x$, the Sinai-Ruelle-Bowen theorem provides
the existence of the steady state measure, denoted
$\mu_{\scriptscriptstyle SRB}$. It can be shown that for a continuous
function $f(x)$, there exists a unique measure
$\mu_{\scriptscriptstyle SRB}$ such that
\begin{equation}
  \lim_{t\rightarrow\infty} \frac{1}{t}\int_0^t dt'
  f(x(t))=\frac
  {\int f(x) d\mu_{\scriptscriptstyle SRB}}
  {\int d\mu_{\scriptscriptstyle SRB}}.
\end{equation}
This measure is known to be fractal for various non-equilibrium
systems, and can be explicitly constructed for certain maps
such as the modified Baker's map\cite{revs}. Here a basis of
fractal Takagi functions has been implemented.

For the Lorentz gas, it was shown numerically under general
conditions, and rigorously proven in the linear response regime, that
the steady state measure is singular\cite{dimensional-loss}. Here one
can explicitly see that the dimensional loss $\Delta D$ is
proportional to the transport coefficient, which in this case is the
electrical conductivity. While there is still some contention
concerning the existence of singular measures in the extension of the
Gibbs ensemble to the far from equilibrium steady state, the
compelling evidence suggests that whatever its nature is, it is far
from trivial.
\section{Transport}
One of the relevant observables is the stress-energy tensor,
\begin{equation}
  {\cal T}^{\mu\nu} = -\frac{\partial {\cal L}}{\partial
    (\partial_\mu\phi)} \partial_\nu\phi 
  + \eta_{\mu\nu} {\cal L}\quad.
\end{equation}
{}From the continuity equation $ 0=\partial_\mu{\cal
  T}^{\mu\nu}$, we have
\begin{eqnarray}
  \label{continuity}
  0 =\frac{\partial}{\partial x^\mu}{\cal T}^{0\mu} =
    \frac{\partial {\cal T}^{00}}{\partial t } + \partder
    {{\cal      T}^{0i}}{x^i}\quad.
\end{eqnarray}
We can identify the heat flux as ${\cal T}^{0i}$.   On the
lattice and in one spatial dimension,
\begin{equation}
  {\cal T}^{01}(x) = -\pi(x)\nabla\phi(x) \rightarrow 
  {\cal     T}^{01}_k=-\pi_k (\phi_{k+1}-\phi_k).
\end{equation}
Defining the lattice energy density consistently as 
\begin{equation}
  \label{t00}
  {\cal T}^{00}(x) = {1\over2}\pi^2+{1\over2}(\nabla\phi)^2
  +V(\phi)\rightarrow 
  {\cal   T}^{00}_k= {1\over2}\pi_k^2+{1\over2}(\phi_{k+1}-\phi_k)^2
  +V(\phi_k), 
\end{equation}
we find that the discrete version of the continuity equation is
satisfied:
\begin{equation}
  \label{cont-disc}
  \partder{}t{\cal   T}^{00}_k+
  \left({\cal   T}^{01}_{k} - {\cal   T}^{01}_{k-1}\right)=0\quad.
\end{equation}
This establishes that ${\cal T}^{01}_k$ is a constant in space and
time in steady state.  On the other hand, there is no way to satisfy
the spatial component of the continuity equation, $\partial_t{{\cal
    T}}^{01}_k+ \nabla{\cal T}^{11}=0$.  The reason for this can be
understood as follows; while the translation invariance in the time
direction is preserved, the translation invariance in the spatial
direction has been lost due to the lattice discretization.

It is interesting to contrast this with the so called FPU
$\beta$ model, which has divergent thermal conductivity in
one spatial dimension. In this case the Hamiltonian is
\begin{equation}\label{fpua}
  H_{\scriptstyle FPU} = \frac{1}{2}\sum_k\left[ p_k^2 +
  (q_{k+1}-q_k)^2 + {\beta\over2} (q_{k+1}-q_k)^4\right],
\end{equation}
so that the heat flux reads
\begin{equation}
 {\cal T}^{01} = -\pi_k (q_{k+1}-q_k) 
 \left(1+ \beta(q_{k+1}-q_k)^2\right).
\end{equation}
If we think of this as originating from a continuum field
theory, we would have the following (non-relativistic)
Lagrangian and heat flux:
\begin{equation}\label{fpub}
  {\cal L} = -\frac{1}{2}
  \left(\partder{\phi}{t}\right)^2
  +{1\over2}\left(\partder\phi x\right)^2
  +\frac{\beta}{4}\left(\partder\phi x\right)^4,
  \qquad {\cal T}^{01}=-\partder \phi t
  \partder\phi   x
  \left(1+\beta\left(\partder\phi x\right)^2\right)\quad.
\end{equation}
It is straightforward to check that the heat current satisfies the
continuity equation (\ref{continuity}). In this case, the thermal
conductivity diverges as $L^{2/5}$ \cite{lepri}.

In the simulations we will define the temperature locally using the
concept of an ideal gas thermometer, which measures the second moment
of the momentum distribution. Specifically, at a site $k\ (x=ka)$ we
define in {\sl both equilibrium and non-equilibrium}
\begin{equation}\label{eq:temp}
 T(x) = T_k = \langle \pi_k^2\rangle.
\end{equation}
We will see that this definition is sensible in these regimes in the
context of both statistical mechanics and thermodynamics.
\subsection{Green-Kubo Approach}
The standard approach to near equilibrium or local equilibrium
transport is to apply linear response theory or Green-Kubo formulas.
In this approach, external fields are added to a Hamiltonian which
generate the desired transport processes and expressions for the
transport coefficients can be derived in the weak field limit.
For thermal conduction, the arguments are somewhat
heuristic since one does not have well defined external fields that
produce heat flow. Nevertheless, the procedure empirically results in
sensible expressions. The Green-Kubo approach expresses transport
coefficients in terms of equilibrium autocorrelation functions. For
the thermal conductivity, one has
\begin{eqnarray}  
  \label{green-kubo}
  \kappa(T)&=&{\frac{1}{T^2}}
  \int_0^\infty\!\!\!\!dt\int\!\!d{x}\,
  \left\langle {\cal T}^{01}({x},t)
    {\cal T}^{01}({x}_0,0)\right\rangle_{\scriptscriptstyle EQ} 
  \nonumber\\
  &=& {\frac{1}{N T^2}}\int_0^\infty\!\!dt\,
  \sum_{k,k'=1}^N  \left\langle {\cal T}^{01}(x_k,t)
    {\cal T}^{01}(x_{k'},0)\right\rangle_{\scriptscriptstyle EQ}.
\end{eqnarray}
$N(<L)$ is the number of sites in the region inside the
boundaries used in the computation.  We typically choose the
region to be as large as possible while excluding the boundary
effects.  While this expression is expected to hold near
equilibrium, the region of its validity cannot be determined
within the linear response theory itself.

Autocorrelation functions, such as Eq. (\ref{green-kubo}),
have been argued to decay
algebraically, rather than exponentially. Originally observed in early 
molecular dynamics simulations\cite{alder}, the velocity-velocity
autocorrelation function was found to decay as
$ \langle v(t) v(0)\rangle_{\scriptscriptstyle EQ}
\sim t^{-d/2}$
up to times on the order of a few ten times the mean free time.
This behavior, if it persists, leads to divergence of the diffusion
constants for $d<3$. Using kinetic theory and some general
assumptions, it was later argued that this is a generic feature of
dynamical systems for long times. These long-time power law tails
cause divergences in transport coefficients in a large class of low
dimensional systems including the FPU $\beta$ model
(\ref{fpua})---(\ref{fpub}).  It is believed, however, that for
theories without strict momentum conservation, such a divergence can
be absent.  This is the case for our lattice model due to our
`on-site' nature of the potential, as we shall see below.
\subsection{Non-Equilibrium Approach}
We also compute the thermal conductivity directly by constructing
ensembles near equilibrium which lead to constant gradient thermal
profiles.  This is arguably more fool-proof than the Green-Kubo
formula since no assumptions are necessary for the computation.  
Here we first confirm Fourier's law and then use it to compute the
conductivity using 
\begin{equation}
  \label{eq:fourier}
  \kappa(T) = -\frac{ \langle {\cal T}^{01}
    \rangle_{NE}}{\nabla T(x)}.
\end{equation}
While Fourier's law is believed to be valid near equilibrium, what
constitutes the `linear regime' is not known. We will see that it
eventually breaks down as we move far from equilibrium, while still in
the steady state.  It is not clear at this point how to make a
sensible definition of the thermal conductivity.  One approach would
be to try to the characterize the `nonlinear' response by
incorporating the dependence of the conductivity on higher order terms
in the derivatives, such as $(\nabla T)^3, (\nabla^2T)(\nabla T)$ and
so on.  We will try to clarify these issues below.
\section{ Equilibrium Ensemble}
When the boundary temperatures are equal, we recover equilibrium
physics for our system.  In the remainder of the paper we specialize
to the $m^2=0$ case.
\subsection{Thermalization}
\begin{figure}
\begin{center}
    \leavevmode
  \epsfxsize=12cm\epsfbox{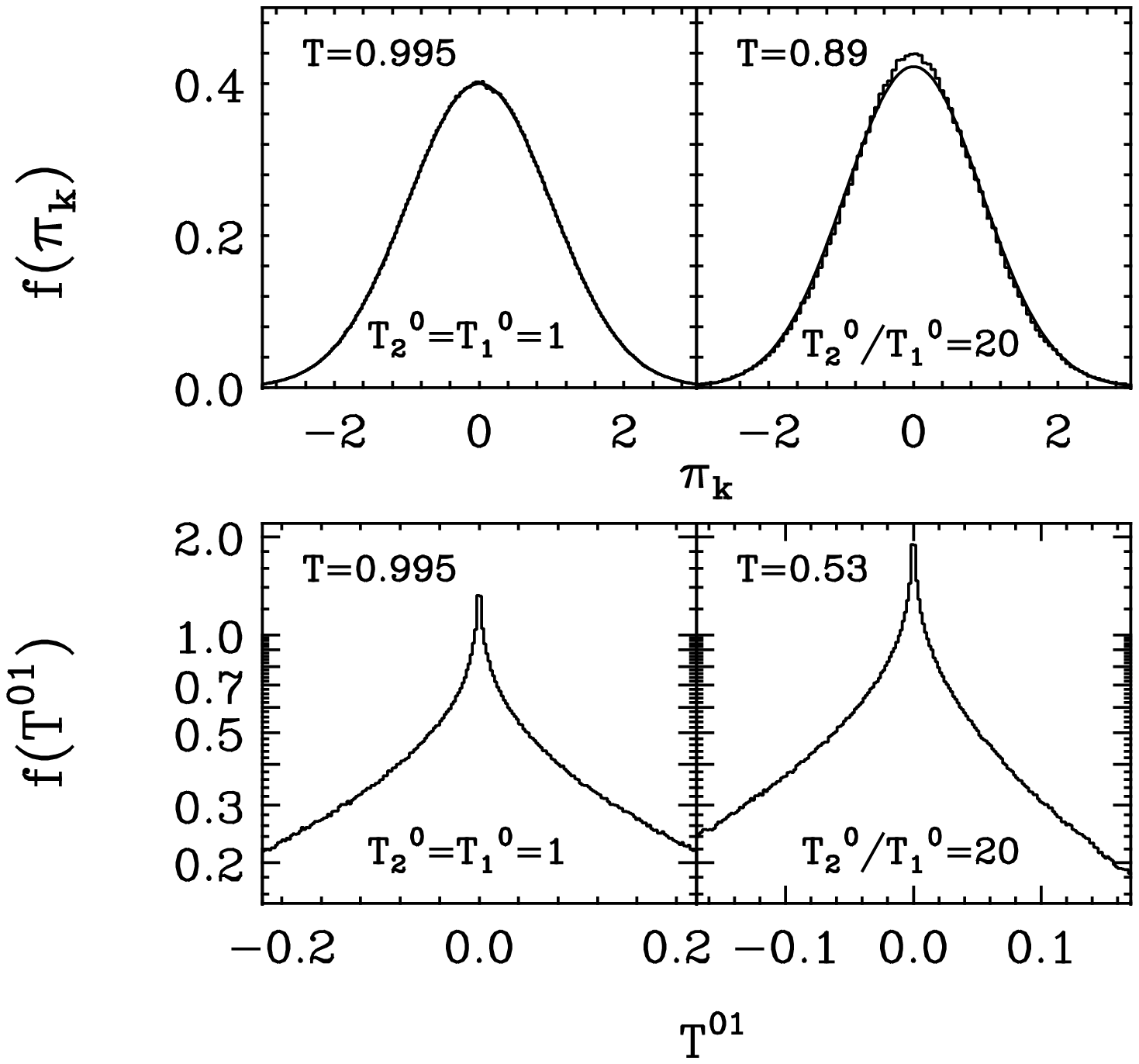}
\caption{Comparison between equilibrium (left)  and non-equilibrium
  (right) steady state distribution functions for momenta $\pi_k$ and
  heat flux ${\cal T}^{01}$. The measured temperature at the site $k$
  is indicated by $T$, and the boundary temperatures by $T_1^0,T_2^0$.
  All measurements are made at mid lattice, $x=k=L/2$, with the
  exception of the lower right figure which was made at $x=L/4$. The
  heat flux $\langle {\cal T}^{01}\rangle$ is zero in equilibrium, so
  $f({\cal T}^{01})$ is symmetric. Away from equilibrium, $\langle
  {\cal T}^{01}\rangle\not =0$, so the distribution develops asymmetry as
  seen in the lower right.}
\label{fig1}
\end{center}
\end{figure}

A test of the thermalization of the system is readily performed by
studying the distribution functions of various quantities. If we take
a single trajectory, $(\phi(t),\pi(t))$, and histogram $\pi_k(t)$ at
$t=0$, $\Delta t$, $2\Delta t$,..., where $\Delta t$ is some time step
which ensures the points are reasonably well decorrelated, we will
converge to the thermal distribution function
\begin{equation}
  f(\pi_k)\sim \exp[-\pi_k^2/2T].
\end{equation}
In Fig. \ref{fig1} (left column), we show computed equilibrium
distributions for the momenta and heat flux at the center of a lattice
with $L=163$. Here we have taken the endpoint temperatures to be
$T_1^0=T_2^0=1$. The measured temperature in the middle is
$T=0.995(8)$, and one can see that the measured momentum distribution
(histogram, top left) agrees with the predicted value (solid, top
left). In the bottom left we show the measured thermal distribution of
${\cal T}^{01}$, for which we do not have a theoretical prediction.
Since we are in equilibrium, there is no heat flow, so $\langle {\cal
  T}^{01}\rangle=0$, which is evident from the symmetric nature of the
thermal distribution $f( {\cal T}^{01})$. 
We have verified that when $T_1^0=T_2^0$, these boundary conditions
reproduce the equilibrium canonical measure $f_{eq}(\pi,\phi)\propto
\exp[-H(\pi,\phi)/T_1]$ at all points, including the thermostatted
sites.  The measured thermal profiles satisfy $T(x)=T^0_1(=T^0_2)$ for
any $x$ within numerical error over the temperature range we
investigated: $T=0.01$ to $T=10$.  Hence the boundary conditions 
(\ref{noneqeom}) do produce the desired physics for this theory.

\begin{figure}
\begin{center}
    \leavevmode
  \epsfxsize=9cm\epsfbox{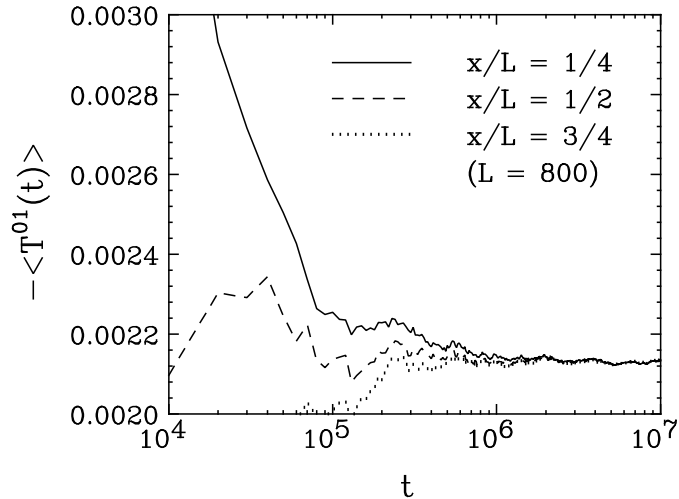}
  \caption{Time evolution of the heat flux $\langle {\cal
      T}^{01}\rangle_{NE}$ at three points inside a system with
    $L=800$, denoted by the ratio $x/L$, and
    $(T_1^0,T_2^0)=(0.3,0.7)$. As there are no sources or sinks
    inside, the asymptotic values must converge to the same value and
    they do.}
\label{fig2}
\end{center}
\end{figure}

In our presentation of equilibrium and non-equilibrium steady state
expectation values, we verify that stationary results are obtained.
This is done by examining the time evolution of observables. For
instance, if we want to measure the heat flow through the system, we
measure at all sites and verify that the time averages converge to the
same value. A typical result is shown in Fig. \ref{fig2}, where the
average $\langle {\cal T}^{01}\rangle$ is shown as a function of time
at three sites: $L/4$, $L/2$ and $3L/4$ for $L=800$. One can see that
the values eventually converge to the ensemble average. In this case
the endpoint temperatures are distinct, so that there is a net heat
flux. Equilibrium figures are similar, with convergence to $\langle
{\cal T}^{01}\rangle=0$.
\subsection{Auto-Correlation Functions}
\begin{figure}
\begin{center}
    \leavevmode
  \epsfxsize=9cm\epsfbox{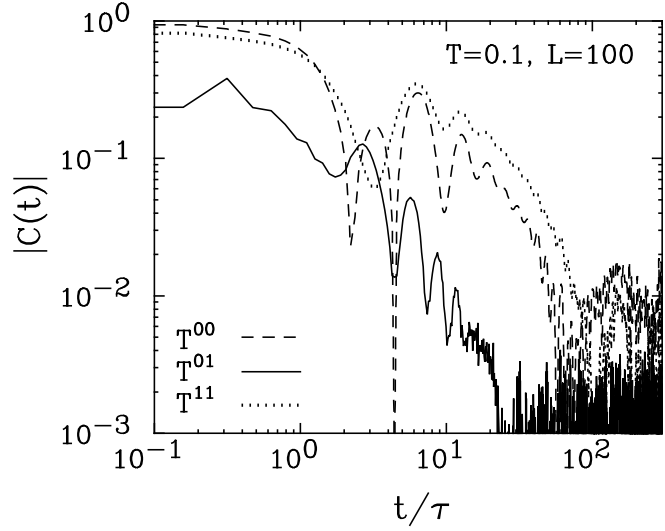}
\caption{Finite temperature autocorrelation functions for ${\cal T}^{\mu\nu}$
  as a function of time scaled by the mean free time
  $\tau\sim\kappa(T)$. We plot the absolute value of $C(t)=(\langle {\cal
    T}^{\mu\nu}(t) {\cal T}^{\mu\nu}(0)\rangle -\langle {\cal
    T}^{\mu\nu}(0)\rangle^2)/ \langle( {\cal
    T}^{\mu\nu}(0))^2\rangle$. At large times, these functions
  oscillate about zero.}
\label{fig3}
\end{center}
\end{figure}

We compute auto-correlation functions for the components of the
stress-energy tensor, which includes the Green-Kubo formula for the
thermal conductivity.  We define the normalized correlation functions
\begin{equation}
 C(t)=\frac{1}{\langle({\cal T}^{\mu\nu}(0))^2\rangle}\left[\langle
            {\cal T}^{\mu\nu}(t) {\cal T}^{\mu\nu}(0)\rangle - 
            \langle {\cal T}^{\mu\nu}(0)\rangle^2\right].
\end{equation}
In Fig. \ref{fig3}, we show the time dependence of $\mid
C(t)\mid$ for ${\cal T}^{00}$ (dashes), ${\cal T}^{01}$ (solid)
and ${\cal T}^{11}$ (dots) as a function of time, where the time
is normalized by the mean free time $\tau$. (As we discuss
below, $\tau$ will be of order of the thermal conductivity.) One
can see that these functions decay to half their initial values on
the order of the mean free time. For larger
times $t\gsim10\tau$, these functions oscillate about zero. This
behavior is seen at all temperatures studied.

\begin{figure}
\begin{center}
    \leavevmode
  \epsfxsize=9cm\epsfbox{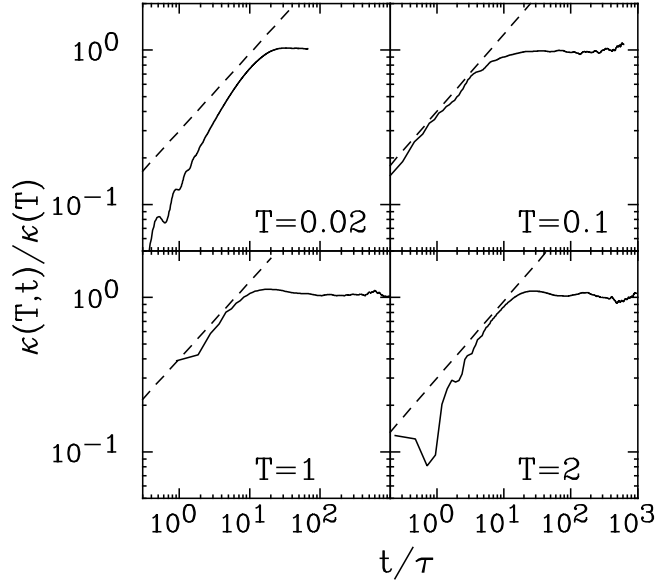}
  \caption{Green-Kubo integrals $\kappa(T,t)$ up to time $t$ for
    lattices with $L=100$. As $t\rightarrow\infty$,
    $\kappa(T,t)\rightarrow \kappa(T)$. We plot the ratio of
    $\kappa(T,t)$ to its asymptotic value versus time, normalized by
    the mean free time $\tau$. Since $C_V\sim c_s\sim 1$, $\tau$ is
    approximated by the magnitude of the conductivity at that
    temperature.  The dashed lines are the anticipated behaviors if
    the long-time tail divergences were present. One can see that on
    the time scales up to $t\sim 10\tau$, such behavior can be seen,
    although it vanishes for larger times.}
\label{fig4}
\end{center}
\end{figure}

To compute the thermal conductivity, we are interested in the
auto-correlation function for ${\cal T}^{01}$:
\begin{equation}
  \kappa(T,t) = \frac{1}{NT^2}  \int_0^t\!\!\!\!dt   
  \sum_{k,k'} \langle {\cal T}^{01}(x_k,t) {\cal
 T}^{01}(x_{k'},0)\rangle_{\scriptscriptstyle EQ},\qquad 
  \kappa(T) = \lim_{t\rightarrow\infty}\kappa(T,t).
\end{equation}
By studying the convergence $\kappa(T,t)$ to the thermal
conductivity, we can explore the problems associated with the
long-time tails.  In Fig. \ref{fig4} we plot $\kappa(T,t)$ as a
function of time, where $t$ is normalized by the mean free time
$\tau$. According to the predictions based on kinetic theory,
in $d=1$ we would expect $\kappa(T,t)\sim t^{1/2}$, leading to an
infinite conductivity.  At the temperatures shown ($T=1/50$, 1/10, 1,
2), $\kappa(T,t)$ does apparently display behavior similar to
${t}^{1/2}$ (dashes) on time scales up to $t\sim 10\tau$, but on
longer time scales, the results converge to well defined values.
Consequently, divergences such as those found in the FPU model are not
present here, and long-time tails are at best a transient aspect of
this dynamics.

\begin{figure}
\begin{center}
    \leavevmode
  \epsfxsize=9cm\epsfbox{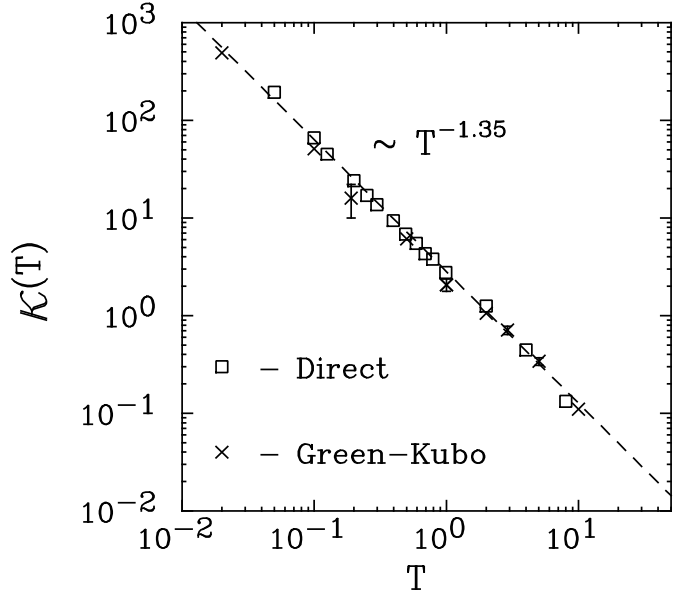}
  \caption{ Power law behavior of the thermal conductivity
  $\kappa$  for lattice $\phi^4$ theory obtained from near
    equilibrium ($\Box$) and Green-Kubo ($\times$) measurements
    for various  lattice sizes $L$. The power law fit
    $\kappa(T)=2.83(4)/T^{1.35(2)}$ is indicated by the dashes.  }
\label{fig5}
\end{center}
\end{figure}

The behavior of $\kappa(T)$ found from the Green-Kubo approach
is summarized in Fig. \ref{fig5} (crosses). This analysis will
be continued when we discuss the direct measurements below.
\subsection{Speed of Sound}
\begin{figure} 
\begin{center}
    \leavevmode
  \epsfysize=8cm\epsfbox{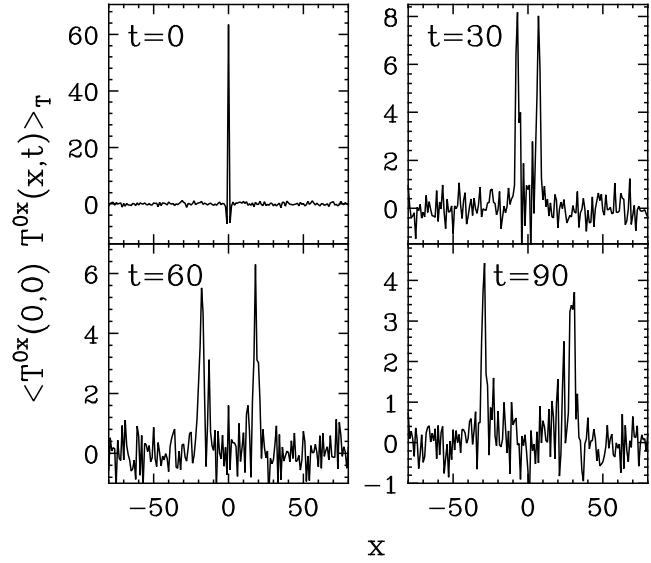}
  \caption{Spatial dependence of the equilibrium
    auto-correlation function 
    $G(x,t;0,0)$ at selected times. $x=0$ is the center of the
    lattice. By examining the peaks, one can estimate how fast the
    excitations propagate through the system.  Here $L=160$ and $T=1/10$. }
\label{fig6}
  \end{center}
\end{figure}

\begin{figure}
\begin{center}
    \leavevmode
  \epsfxsize=9cm\epsfbox{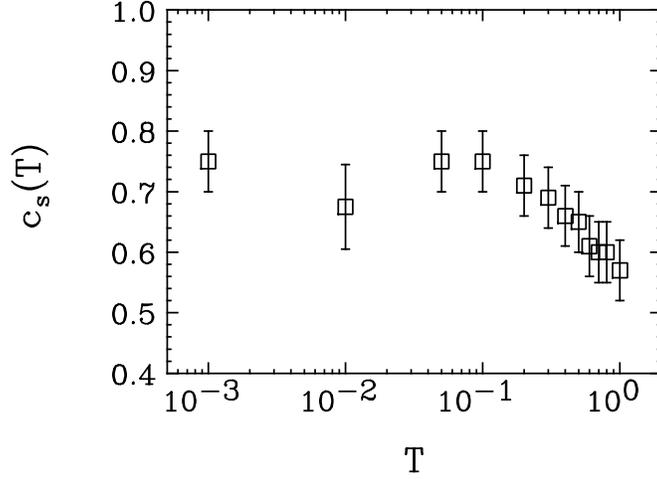}
  \caption{ Temperature dependence of the ``speed of sound'', $c_s$.
    The measurements were made in a system with $L=163$ but the speed
    was also seen to be independent of the size.}
\label{fig7}
\end{center}
\end{figure}

To better understand the kinetic theory aspects of this finite
temperature theory, we would like to know the thermal behavior of the
`speed of sound'. This can be estimated in a number of ways. A
convenient approach is to use the ${\cal T}^{01}$ auto-correlation
function.  The sound speed, $c_s$, is defined here as how fast
excitations travel through the system and is the relevant velocity for
the transport theory in our particular model.  We note, however, that
strictly speaking, this is not `sound' in the hydrodynamic sense. We
define
\begin{equation}
  \label{autoJ}
  G(x,t;x_0,t_0)\equiv\left\langle {\cal
      T}^{01}(x_0,t_0){\cal T}^{01}(x,t)\right\rangle_{EQ}.
\end{equation}
Instead of the time dependence, we consider the spatial dependence of
$G(x,t;x_0,t_0)$. Not only will it decorrelate in time, as we saw
previously in the Green-Kubo integrals, but it will also decorrelate
over space.  A typical behavior of $G(x,t;x_0,t_0)$ is shown in Fig.
\ref{fig6}. Here we choose $x_0=0$ to be the center of the lattice,
and $t_0=0$. The temperature is $T=1/10$ and $L=160$. The
autocorrelation function is shown for times $t=0,30,60,90$ where one
can see the regions of high correlation separate. By measuring the
rate at which the peaks separate, we can obtain an estimate for $c_s$
at that temperature.  In Fig. \ref{fig7}, we plot the
sound speed extracted in this manner against the temperature.  There
is little temperature dependence until $T\sim 1/10$, after which the
speed begins to decrease with $T$, as one might naively expect. So
although it shows some temperature dependence, $c_s$ is generally of
order unity over this temperature range.
\section{ Non-Equilibrium Steady States}
\label{sec:non-eq}
By controlling the boundary temperatures, we can begin to develop an
understanding of the physics of hot scalar field theory, as it moves
increasingly further from thermal equilibrium\cite{plb}. 
We discuss the regimes 
we have found as we move away from equilibrium.

\subsection{Near Equilibrium: $T_1\sim T_2$}

\begin{figure}
\begin{center}
  \leavevmode \epsfxsize=9cm\epsfbox{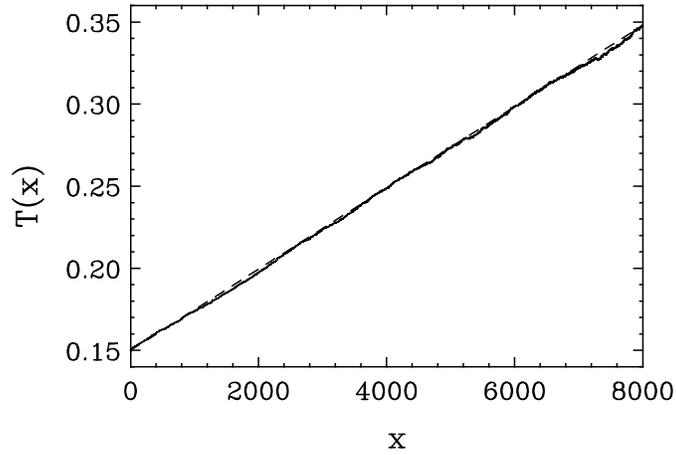}
  \caption{A typical near equilibrium temperature profile
    (solid), for $L=8000$, compared to a linear profile (dashes). The
    two endpoint thermostats are $T_1^0=0.15$ and $T^0_2=0.35$.}
\label{fig8}
\end{center}
\end{figure}

\begin{figure}
\begin{center}
    \leavevmode
  \epsfxsize=9cm\epsfbox{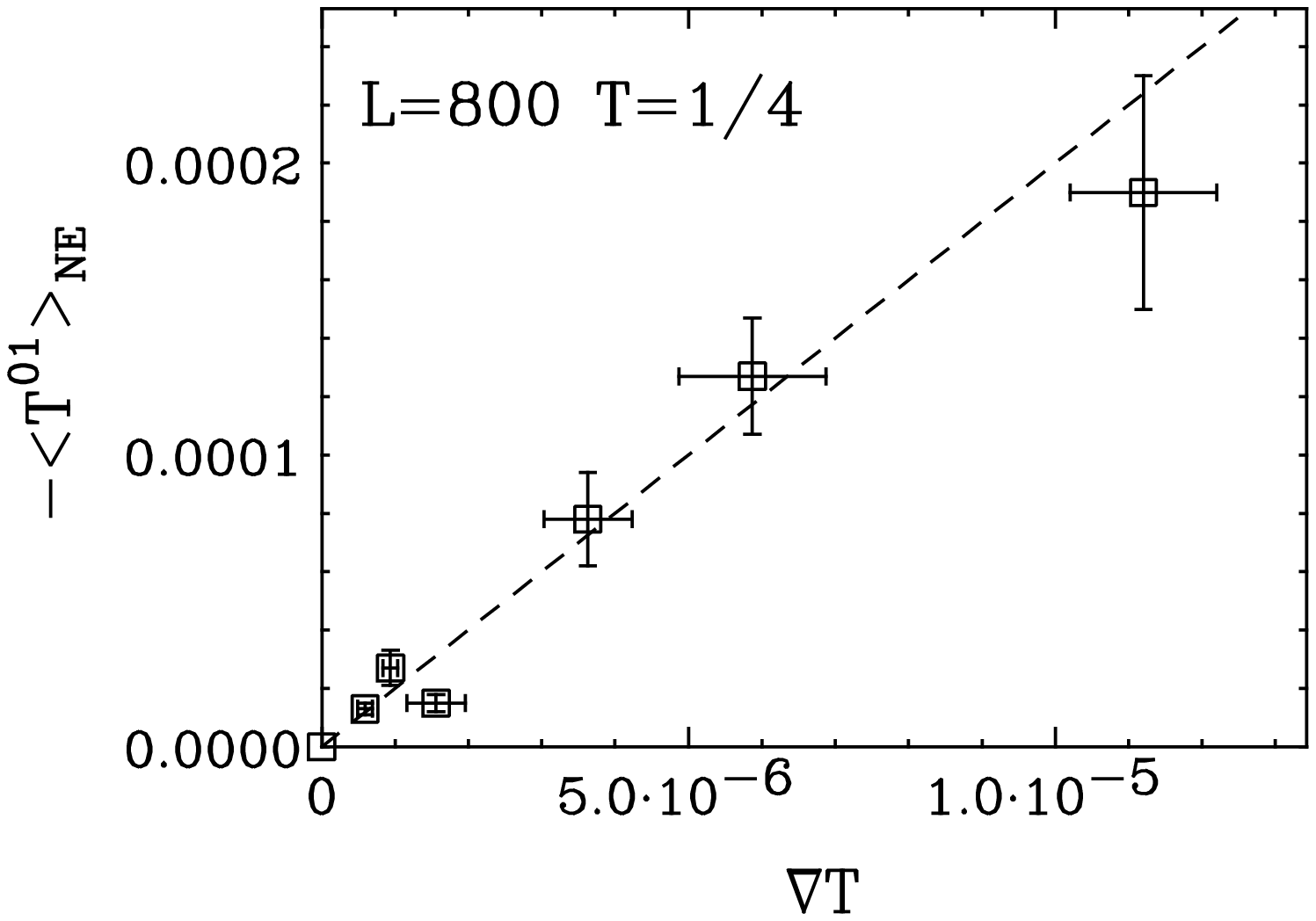}
  \caption{Method used for direct computation of the thermal
    conductivity near equilibrium. In this regime the thermal profiles
    are linear and Fourier's law is well reproduced.  By taking
    increasingly small differences $T_2^0-T_1^0$ around a fixed
    temperature $T$, one can measure $\langle {\cal
      T}^{01}\rangle_{NE}$ for many values of $\nabla T$ and extract
    the slope, which equals the conductivity. Here we show such a
    result for $L=800$ at $T=1/4$.}
\label{fig9}
\end{center}
\end{figure}

In the {\sl near equilibrium} limit, the physics is consistent with
linear response and Fourier's law. Here $\langle {\cal T}^{01}\rangle$
will develop a non-zero expectation value. As the temperatures $T_1$
and $T_2$ begin to separate, a constant $\nabla T$ develops. This is
illustrated in Fig. \ref{fig8} for a lattice of $L=8000$ where a small
gradient is developed around a temperature $T=1/4$ using boundary
temperatures $T_1^0=0.15$ and $T_2^0=0.35$ (solid). A linear profile,
expected from Fourier's law, is superimposed (dashes). While such
linear profiles in principle provide a direct measure of the
conductivity by dividing the measured heat flux by the average
gradient, we perform a more careful analysis by taking increasingly
smaller gradients about the same average temperature, thereby
explicitly verifying Fourier's law. This is shown in Fig. \ref{fig9}
for the temperature $T=1/4$. All these points correspond to thermal
profiles which are linear. As the boundary temperatures approach
$T=1/4$, we see that the heat flux also decreases. The slope,
$\kappa(T)=-\langle {\cal T}^{01}\rangle/\nabla T$, corresponds to the
thermal conductivity at this temperature.

The measurements of $\kappa(T)$ in the range $1/100<T<20$ are
summarized in Fig. \ref{fig5} for both direct measurements ($\Box$)
and Green-Kubo ($\times$). We find that both methods agree and that
the general behavior is well described by a power law:
\begin{equation}\label{kappa}
  \kappa(T)=\frac{A}{T^{\gamma}},\qquad\qquad \gamma=1.35(2),\qquad 
  A=2.83(4).
\end{equation}
This type of behavior is reminiscent of lattice phonons at high
temperature\cite{herring}. The fact that the direct measurements agree
with the Green-Kubo integrals of the auto-correlation functions is
sufficient to dispel the notion of asymptotic long-time tails in this
system.

\begin{figure}
\begin{center}
    \leavevmode
  \epsfxsize=9cm\epsfbox{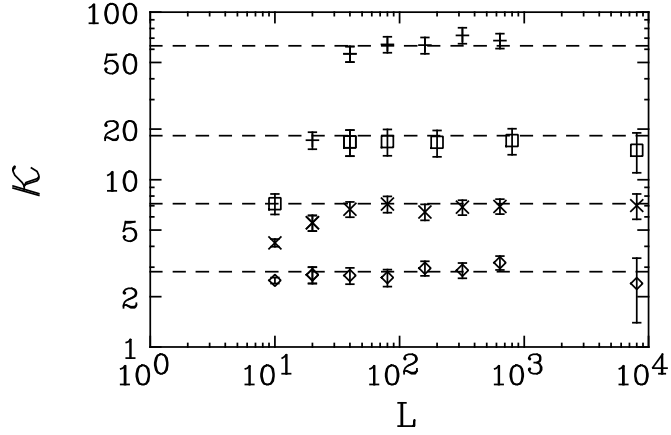}
  \caption{ Evidence of bulk behavior of the thermal conductivity for
    selected temperatures $T=1$ ($\Diamond$), $T=1/2$ ($\times$),
    $T=1/4$ ($\Box$) and $T=1/10$ ($+$). The dashed lines are the
    predictions from the power law fit~(\ref{kappa}) to $\kappa(T)$.}
\label{fig10}
\end{center}
\end{figure}

Another important aspect to verify is that we have achieved a bulk
limit in the values of $\kappa(T)$ reported. In Fig. \ref{fig10}, we
show the dependence of the direct measurements of $\kappa$ on the size
$L$ of the system for several temperatures. The dashed lines are the
predictions from the power law fit. One can see that a bulk limit is
realized for relatively small lattices.  On closer inspection, we find
that the bulk limit is reached for smaller lattices when the
temperature is higher.  This can be understood as follows; the mean
free path of the system is of the order of the thermal conductivity,
as we shall see when we analyze the kinetic theory aspects of this
problem.  The bulk limit is reached for lattices larger than the mean
free path, roughly speaking.
\subsection{Far from Equilibrium: $T_1\ll T_2:$ }
\begin{figure}
\begin{center}
    \leavevmode
  \epsfxsize=9cm\epsfbox{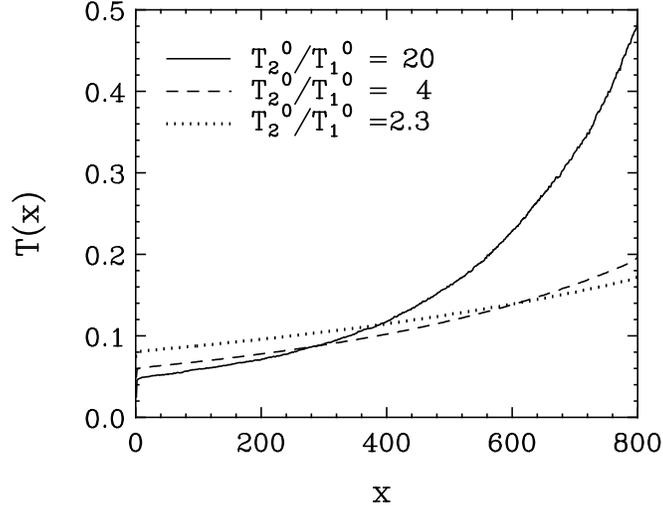}
  \caption{As the boundary temperature differences increase, the
    system departs from the constant gradient regime and the thermal
    profiles develop curvature. In the figure we show, the boundary
    temperatures are $(T_1^0,T_2^0)=(0.3,0.7)$ (dots), $(0.2,0.8)$
    (dashes) and $(0.1,2.0)$ (solid) for a system with $L=800$.}
\label{fig11}
\end{center}  
\end{figure}

We now consider the behavior of the system as we move it further from
equilibrium.  One of the first characteristics to emerge is the
development of curvature in the temperature profile. In
Fig.~\ref{fig11} we plot a succession of steady state temperature
profiles as we change the boundary temperatures from
$(T_1^0,T_2^0)=(0.3,0.7)$ (dots) to $(0.2,0.8)$ (dashes) and finally
to $(0.1,2)$ (solid). As the system moves away from the constant
gradient profiles, it starts to feel the temperature dependence of the
thermal conductivity, which decreases with increasing temperatures. As
a consequence, the hot end of the system cannot conduct heat as well
as to lower temperature end.  (Of course the converse of this argument
would hold if the power law behavior has $\gamma<0$, in which case the
curvature would have an opposite sign.)

Fig. \ref{fig2}, which was already mentioned when we discussed the
convergence properties, displays the time evolution of the heat flux
at three sites within a system thermostatted at
$(T_1^0,T_2^0)=(0.3,0.7)$ (dots in Fig.~\ref{fig11}). Regardless of
how far the system is from equilibrium, the heat flux must be
independent of position in steady state, since there are no sources or
sinks for heat inside the boundaries. We see that the values at the
three sites converges to the same value in the steady state, namely, a
constant non-equilibrium heat flux independent of $x$.

\begin{figure}
\begin{center}
    \leavevmode
  \epsfxsize=12cm\epsfbox{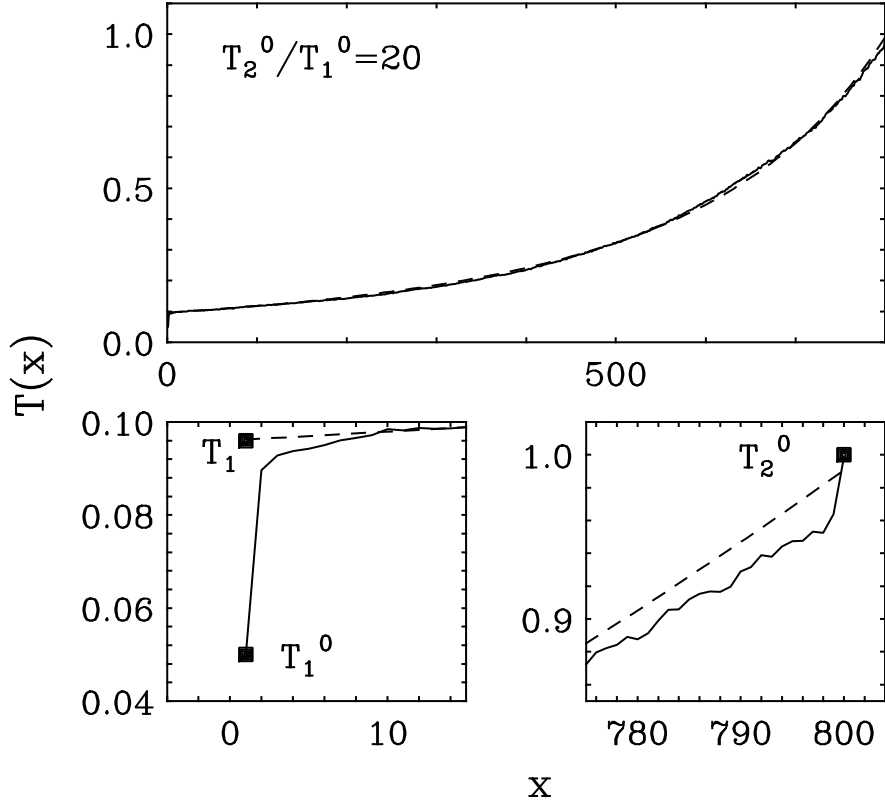}
  \caption{(Top) Non-equilibrium thermal profile for boundary
    temperatures $(T_1^0,T_2^0)=(0.05,1)$ (solid) compared to
    predictions (solid). One can see deviations on the of a few~\%,
    most notably at the high temperature end (bottom, right). The
    boundary jumps are about equal and are readily understood.}
\label{fig12}
\end{center}
\end{figure}

Another property of the non-equilibrium system is the appearance of
jumps in the temperature at the boundaries. This is illustrated in
Fig.~\ref{fig12}; in Fig.~\ref{fig12} (top), we show a typical
non-equilibrium thermal profile with curvature (solid --- the dashes
will be discussed below). In the lower panels we examine the low and
high temperature ends. $T_1^0$ and $T_2^0$ are the temperatures
enforced by our boundary temperatures. One can see that there is a
difference between these temperatures and what one obtains by smoothly
extrapolating the temperature profile near the edge. These are
physical phenomena associated with the dynamics at the interface,
which can be readily understood quantitatively, as discussed in
section~\ref{sec:non-eq}~D below.  We will define the points obtained
through smooth extrapolation by $(T_1,T_2)$. If we now focus on the
thermal profile away from the edges, we can understand the curvature
in the temperature profile using Fourier's law. At or near
equilibrium, we have found the power law behavior $\kappa(T)=A
T^{-\gamma}$, as in Eq.~(\ref{kappa}). If we integrate Fourier's law
and re-express the result in terms of extrapolated temperatures
$(T_1,T_2)$, we find
\begin{equation} \label{t-profile}
  T(x) = \left\{\begin{array}{ll}
    T_1\left[1-\left(1-\left(\frac{T_2}{T_1} \right)^{1-\gamma}
      \right)
      {\frac{x}{L}}\right]^{{\frac{1}{1-\gamma}}},\qquad &
    \gamma\neq1\\ 
    T_1 \left( \frac{T_2}{T_1}\right)^{x/L}, &\gamma=1\quad.
\end{array}\right.
\end{equation}
Some previous efforts to model the temperature profiles in 
systems exist
\cite{tprofs}, although these were generally model fits and
furthermore, a full understanding requires a description of the
boundary effects, which up to now have been lacking.  We have found
that this formula provides an excellent description, but ultimately
breaks down as we see below. In Fig.  \ref{fig12}, the dashed line is
the description given by Eq.  (\ref{t-profile}) with $\gamma=1.35$,
which offers a fit to within a few percent. In should be noted that
$\gamma$ used here was obtained from systems at or near equilibrium as
in Eq.~(\ref{kappa}), independently of the systems far from
equilibrium. One can see that at the high temperature end there is a
tendency to overshoot the measured behavior as in
Fig.~\ref{fig12}~(bottom right).  This analytic expression provides a
sufficiently good description of the physics as we move away from
equilibrium, until Fourier's law fails to hold locally. This includes
the first three regimes listed in Table 1.  Eventually, Fourier's law
will break down locally as well, as we develop steady states which are
locally non-equilibrium (LNE).

The analytic understanding of the thermal profile allows us to
understand the behavior of the heat flux for systems not too far from
equilibrium.  Using Fourier's law and Eq. (\ref{t-profile}), we have
for
\begin{equation}
  \label{ne-j}
  \langle {\cal T}^{01}\rangle_{NE}
 = \frac{A}{L(1-\gamma)}(T_1^{1-\gamma}-T_2^{1-\gamma})
 \qquad(\gamma\not=1).
\end{equation}
For $\gamma=1$, $  \langle {\cal T}^{01}\rangle_{NE}
 = ({A}/{L})\log(T_1/T_2)$.
When compared to measured heat flux using the extrapolated
temperatures from the thermal profile, this formula is found to
provide a very good description of the heat flux, typically to within
a few percent. In the near equilibrium regime where the temperature
profile is visibly linear, the boundary jumps vanish, and we expect it
to behave as
\begin{equation}\label{ne-j1}
  \langle {\cal T}^{01}\rangle_{NE}^0
 = -\kappa(T_{av})\frac{T_2-T_1}{L}
\end{equation}
where the superscript denotes the linear response, or constant
gradient limit. In general, these are related by
\begin{equation}
 \langle  {\cal T}^{01}\rangle_{NE} =  \langle  {\cal
       T}^{01}\rangle_{NE}^0 
       \left\{ 1+\frac{\gamma(\gamma+1)}{24}
      \left(\frac{\Delta T}{T_{av}}\right)^2 + 
      {\cal O}\left(\left[\frac{\Delta
     T}{T_{av}}\right]^4\right)\right\},
\end{equation}
where $T_{av}=(T_2+T_1)/2$ and $\Delta T=(T_2-T_1)$. Note that these
temperatures are the extrapolated ones and not the boundary
temperatures, $(T_1^0,T_2^0)$.
In this
way we see that the temperature profile will no longer be linear when 
${\Delta T}/{T_{av}}\sim L{\nabla T(x)}/{T(x)}\sim 1$.

In this regime of curved thermal profiles, we see some differences
between the thermal distributions measured in equilibrium. In Fig.
\ref{fig1} we display on the right side the measured steady-state
non-equilibrium statistical distributions for momenta (histogram,
upper right) compared to the equilibrium thermal distribution expected
at that temperature based on the notion of local equilibrium (solid).
We find that as we more further from equilibrium, the momentum
distributions are typically sharper than the gaussian expected based
on the ideal gas thermometer.  In the lower right, we show the
non-equilibrium distribution of heat flux which develops the asymmetry
needed to give a non-vanishing expectation value $\langle{\cal
  T}^{01}\rangle<0$ for $T_1^0<T_2^0$.

\begin{figure}
\begin{center}
    \leavevmode
  \epsfxsize=9cm\epsfbox{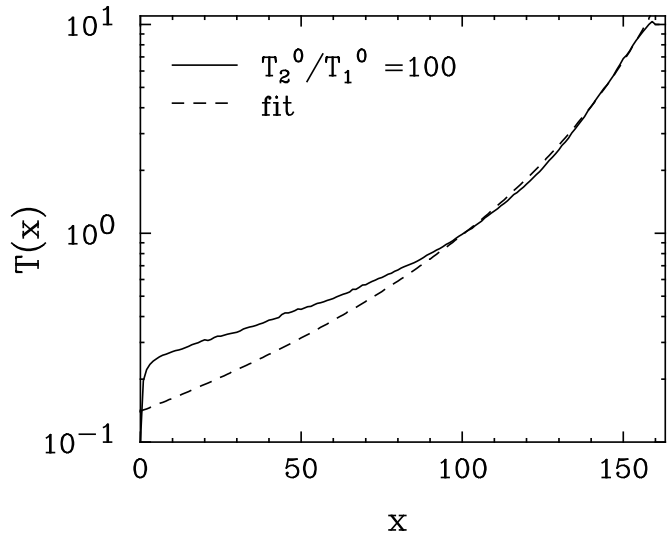}
  \caption{As the system is driven even further from equilibrium, we
    begin to see departures from local equilibrium and linear
    response. Here we show the thermal profile for a system with
    $T_2^0/T_1^0=100$ and $L=163$ (solid), compared to the theory
    curve from Eq. (\ref{t-profile}) (dashes). The agreement is not
    good.}
\label{fig13}
\end{center}
\end{figure}

\begin{figure}
\begin{center}
    \leavevmode
  \epsfxsize=9cm\epsfbox{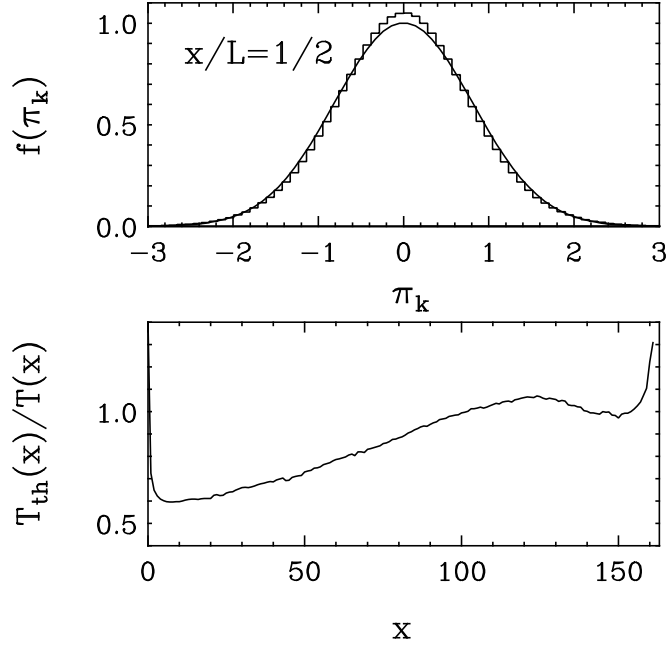}
  \caption{(Top) Steady state distribution of momenta $\pi_k$ in
    the center of the strongly non-equilibrium system discussed in the
    previous figure. We observe that these distributions are
    characteristically more strongly peaked than the local equilibrium
    gaussian at that temperature (solid). (Bottom) The ratio of the
    theory profile to the measured profile in the previous figure, as
    a function of position. The agreement is worse for lower
    temperatures.}
\label{fig14}
\end{center}
\end{figure}

An illustration of a system very far from equilibrium is the
non-equilibrium steady state shown in Fig. \ref{fig13}. Here the ratio
of the endpoint temperatures is 100, and the thermal profile is shown
on a log scale (solid). The fit of Eq. (\ref{t-profile}) is shown by
the dashed line. While we still measure temperature locally through
the second moment, the measured distributions are no longer gaussian.
In Fig. \ref{fig14} (top) we show $f(\pi)$ in the center of the
system. In the bottom panel, the ratio of Eq. (\ref{t-profile}) to the
measured behavior in Fig. \ref{fig13} is shown. Deviations up to 40\%
are evident on the low temperature end. Here, there is no accepted
formalism to describe the dynamics and we are in the LNE regime of
Table 1.  We will return to this issue when we discuss local
equilibrium in more detail in section~\ref{sec:non-eq}~F.
\subsection{Entropy and Kinetic Theory} 
While the Gibbs entropy for the theory diverges as it contracts onto a
set of measure zero, not all measures of entropy behave in this
manner. Irreversible thermodynamics provides a description of systems
on scales much larger than the mean free path\cite{irrev}. When one
has heat flow, one defines the local rate of entropy production as
\begin{eqnarray}
 \sigma(x) &=& \langle  {\cal T}^{01}\rangle_{NE}\frac{d}{dx}
 \frac{1}{T(x)}\\ 
 &=& \frac{A}{L^2 (1-\gamma)^2 T_1^\gamma} 
      \left[ 1-\left(\frac{T_2}{T_1}\right)^{1-\gamma}\right]^2
       \left[1-\left(1-\left(\frac{T_2}{T_1}
      \right)^{1-\gamma} \right)
    {\frac{x}{L}}\right]^{(2-\gamma)\over(\gamma-1)}>0\quad.\nonumber
\end{eqnarray}
Integrating this formula, we can compute the net rate of entropy
production to be
\begin{eqnarray}
  \dot S_{irr}
  &=&\int_0^L dx\; \sigma(x)\nonumber\\
  &=& \langle {\cal T}^{01}\rangle_{NE}\left
    ( \frac{1}{T_2}-\frac{1}{T_1}\right)\\
  &=& \frac{A}{L(\gamma-1) T_1^\gamma}
  \left(1-\frac{T_1}{T_2}\right)
  \left[1-\left(\frac{T_1}{T_2}\right)^{\gamma-1} \right]  
  >0.\nonumber
\end{eqnarray}
This expression can be simply interpreted as saying that the global
entropy production rate is due to the difference in entropy production
rates at the boundaries coming from the demons. On the other hand,
irreversible thermodynamics predicts a constant local rate of entropy
production, $\sigma(x)$, which is at odds with the coarse-grained local
Boltzmann 
entropy computed below. The latter calculation is microscopically
based and does not rely on any hydrodynamic limit of the theory.

{}If we envision the expansion of hot systems similar to RHIC
collisions, the local entropy is an important quantity. From the
statistical mechanics point of view, when one discusses the notion of
entropy in a local frame, it is not Gibbs entropy of the entire
systems that is needed. Hence we would like to consider Boltzmann's
entropy.  To understand the behavior of this entropy in increasingly
non-equilibrium environments, we consider the $n$-body Boltzmann
entropy, defined through the $n$-body distribution functions
$f^{(n)}_B$. The best we can do, of course, is the coarse-grained
limit of these quantities. In the `local frame' at $x=x'$, we define
them as integrals of the full phase space distribution $f$ over all
quantities except the arguments of $f_B^{(n)}$:
\begin{eqnarray}
 f^{(1)}_B(\phi(x'),\pi(x'))&=&f^{(1)}_B(\phi_k,\pi_k)\nonumber\\
 f^{(2)}_B(\phi(x'),\phi(x''),\pi(x'),\pi(x''))&=&
    f^{(2)}_B(\phi_k,\phi_{k+1},\pi_k,\pi_{k+1})\\
  &\vdots &.\nonumber
\end{eqnarray}
The distributions $f_B^{(k)}$ are obtained  by histogramming the
corresponding degrees of freedom ( $(\phi_k,\pi_k)$ for $f^{(1)}_B$, 
$(\phi_k,\pi_k,\phi_{k+1},\pi_{k+1})$ for $f^{(2)}_B$,...), and is 
readily constructed in equilibrium and non-equilibrium conditions.
In the equilibrium or non-equilibrium steady states, we then compute
\begin{eqnarray}
  S^{(1)}_B &=& -\int d\pi(x) d\phi(x) \, 
  f^{(1)}_B\log f^{(1)}_B  \nonumber\\
  S^{(2)}_B &=& -\int d\pi(x) d\phi(x)   d\pi(x') d\phi(x') 
          \, f^{(2)}_B\log f^{(2)}_B\\
        & \vdots& \nonumber
\end{eqnarray}
These satisfy the inequalities
\begin{equation}
  S_G
  \leq \cdots\leq \frac{L}{2} S^{(2)}_B \leq L S^{(1)}_B.
\end{equation}
To compute $S_B$ we must coarse-grain the phase space. Hence we are
actually computing the coarse-grained 1- and 2-body densities, which
we denote $f_\Delta^{(k)}$. These are related by
\begin{equation}
   S^{(1)}_B \simeq -\sum f_\Delta \log f_\Delta - \log
   (\Delta\pi_k\Delta\phi_k),\qquad\sum f_\Delta = 1.
\end{equation}
We have computed these entropies and find that $S_B^{(1)}$ does not
shift noticeably from its equilibrium value regardless of how far the
system is from equilibrium (see Fig.~\ref{fig15}). Further, $S_B^{(2)}$
$(\alt 2 S_B^{(1)})$ is only slightly less than its upper limit $2
S_B^{(1)}$ and remains so even far from equilibrium.  So unlike $S_G$
$\left(\leq L S^{(1)}_B\right)$, $S_B$ is rather insensitive to the
non-equilibrium nature of the system.  Since this is coarse-grained,
it is not too surprising that this is so. It would be more revealing
to consider the behavior of $f_B^{(k)}$ as the size of the bins
decreases, but this is currently too computationally intensive.

\begin{figure}
\begin{center}
    \leavevmode
  \epsfxsize=12cm\epsfbox{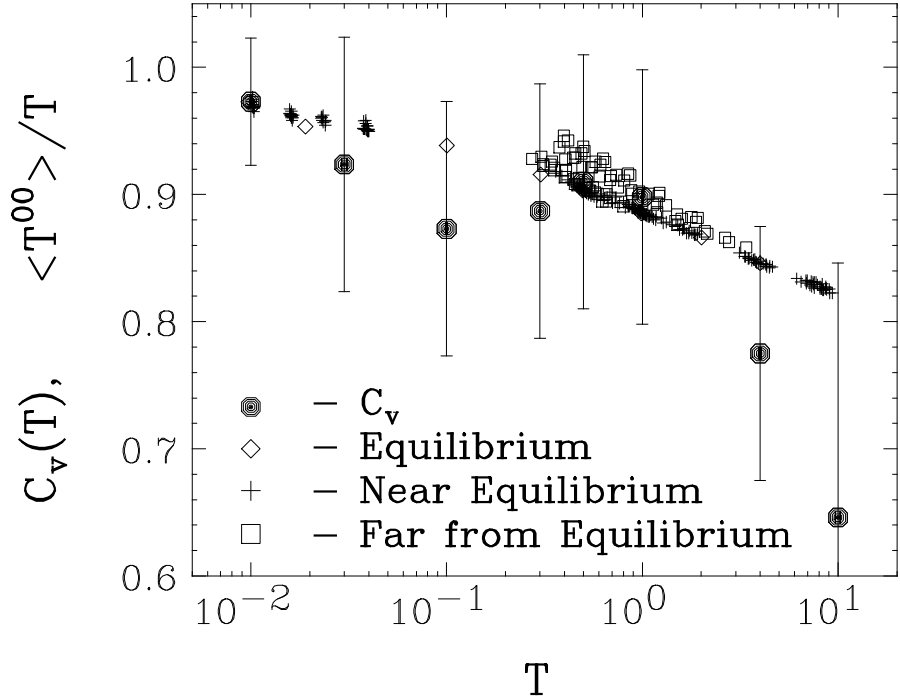}
  \caption{Specific heat, $C_V$, measured in equilibrium is
    plotted as a function of temperature.  The ratio of the energy
    density $\langle {\cal T}^{00}\rangle$ is also plotted against the
    temperature $T(x)$, at equilibrium ($\Diamond$), near equilibrium
    ($+$) and far from equilibrium ($\Box$).  We see that there is a
    temperature dependence and as well as seemingly a slight effect
    due to non-equilibrium physics.
    }
\label{fig16}
\end{center}
\end{figure}

\begin{figure}
\begin{center}
    \leavevmode
    \epsfxsize=12cm\epsfbox{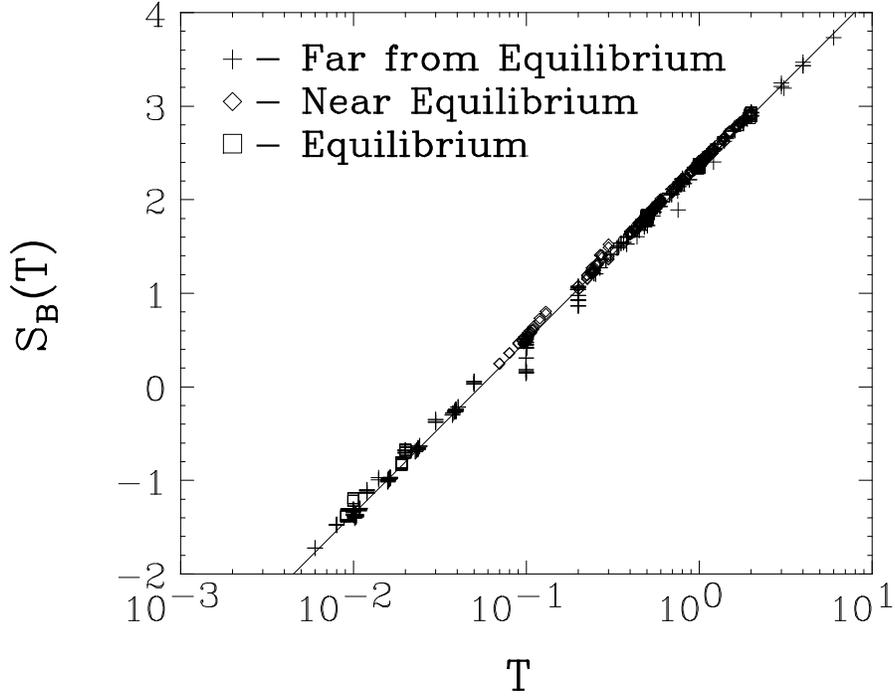}
  \caption{Coarse-grained one-body Boltzmann entropy $S_B(T)$ as
    a function of $T$ for equilibrium ($\Diamond$), near equilibrium
    ($+$) and far from equilibrium ($\Box$) systems. One can see that
    the (local) entropy is insensitive to the departure from
    equilibrium. The solid line indicates a logarithmic
    fit, $S_B(T)=S_{B,0}+S_{B,1}\log T$, with $S_{B,0}=2.5(1)$
    and $S_{B,1}=0.80(2)$.
    }
\label{fig15}
\end{center}
\end{figure}

Thermodynamic quantities, such as the entropy, allow us to investigate
the underlying dynamics of the theory and also probe the possible
deviations of the physical observables from their equilibrium values
under non-equilibrium conditions.  We shall analyze these questions
below.  Let us first obtain the specific heat, $C_V$, from the
equilibrium ensembles; $C_V$ may be obtained using the standard
formula,
\begin{equation}
  \label{cv-eq}
  C_V={\langle E^2\rangle_{EQ}-\langle E\rangle_{EQ}^2\over T^2} 
  , 
\end{equation}
where $E$ is the energy per site. We find that the specific heat
has a {\it weak} temperature dependence which may be fitted by a
simple temperature dependence,
\begin{equation}
  \label{cv-t-dep}
  C_V=C_0 T^{-\alpha},\qquad C_0=0.86(2),\qquad
  \alpha=0.025(6) \quad.
\end{equation}
This entails that the energy of the lattice per site has the
following behavior
\begin{equation}
  \label{E-t-dep}
  E={C_0\over1-\alpha}T^{1-\alpha}\quad
  \arra  \quad E=C_0 T\quad.
\end{equation}
Since the temperature dependence is weak, it is natural to compare
$C_V$ against $\langle E\rangle/T$ obtained {\it locally} both in
equilibrium and in {\it non-equilibrium}.  Such a comparison of $C_V$,
$\langle E\rangle/T$ for equilibrium, near equilibrium and far from
equilibrium is shown in Fig.~\ref{fig16}.  We find that they all agree
quite well.  There seems to be an intriguing tendency for the
non-equilibrium values of $\langle E\rangle/T$ to be larger than their
equilibrium counterparts.  However, this cannot be delineated
within the errors and more investigation is necessary to see if this
is a real effect.

Let us now move onto the computation of the {\it Gibbs entropy},
$S_G$, {\it in equilibrium}.  From the application of the first law of
thermodynamics, we obtain
\begin{equation}
  S_G = \int^T\!\!\frac{C_V}{T}dT.
\end{equation}
\begin{equation}
  \label{eq:ent}
  S_G(T) = S_{G,0} 
  - {C_0\over\alpha}\left(T^{-\alpha}-1\right) 
  \quad\arra\quad S_{G,0}  + C_0\ln T\quad.
\end{equation}
In Fig.~\ref{fig15} we plot the computed {\it Boltzmann entropy}
$S^{(1)}_B$ as a function of temperature for systems in equilibrium,
as well as for systems near and far from equilibrium and find a
similar form, $S_B^{(1)}=S_{B,0}+S_{B,1}\log T$.  While we cannot
measure $S_G$ away from equilibrium, $S_{B,1}\simeq C_0$ so that the
behavior of $S_B^{(1)}$ in equilibrium and in {\it non-equilibrium} is
similar to that of $S_G$ {\it in equilibrium}.  
We also find that $S^{(1)}_B$ is completely insensitive to the
departure from equilibrium.  Consequently, Boltzmann's entropy and
some of the thermodynamic concepts such as temperature, still retain
their relationships in these local states far from equilibrium, even
when the momentum distributions are no longer gaussian.
Other physical quantities were also studied for effects of the system
being in non-equilibrium, including ${\cal T}^{11}$,
\begin{equation}
  \label{t11}
  {\cal T}^{11}= {1\over2}\pi^2+{1\over2}\left(\nabla\phi\right)^2 
  -{\phi^4\over4}\quad.
\end{equation}
We find that the results for ${\cal T}^{11}$ in equilibrium, near
equilibrium and far from 
equilibrium are consistent with each other in a manner similar to
${\cal T}^{00}$ in Fig.~\ref{fig16}.
\subsection{Boundary Temperature Jumps}
Boundary temperature jumps in non-equilibrium steady states, such as
systems experiencing thermal gradients are well
known\cite{jump-sim,hatano}, although it seems that their behavior has
never had a suitable explanation.  Systems experiencing shearing due
to moving walls also display jumps in the velocity profile, with the
fluid inside the wall moving slightly slower than the wall velocity,
for large velocities.  Such effects are known to be sources of error
in experiments that measure transport coefficients \cite{exp}. We find
that it is possible to achieve a quantitative understanding of these
effects using simple kinetic arguments.

We have found in this system that $c_s$ and $C_V$ are of order unity,
and have at most a weak temperature dependence (see Fig.~\ref{fig7}\ 
and Fig.~\ref{fig16}). The thermal conductivity is related to the mean
free path by $\kappa\simeq C_V c_s \ell$ by a standard kinetic theory
argument. Hence we expect $\kappa\sim \ell$.  Strictly speaking,
$\ell$ is the mean free path {\it in equilibrium}, but we will refer
to this loosely as the mean free path also away from equilibrium, to
use as a natural length scale in the system.  It corresponds to the
mean free path when the thermal gradients are not too strong (up to
LE-I$\!$I regime in Table 1), but the standard kinetic theory
arguments presumably break down when local equilibrium no longer
holds.  The boundary temperature jumps are due to the mean free path
being non-zero and the deviation of the system from equilibrium. When
$\ell\ll L$,
\begin{equation}\label{eq:b-jump}
  T_i^0-T_i=\eta \left.\frac{\partial T}
   {\partial n}\right|_{boundary} 
 ,
\end{equation}
where $n$ denotes the normal to the boundary \cite{pk}.  This formula
should apply when the jump $|T_i^0-T_i|/T_i$ is relatively small.
On dimensional grounds, the coefficient $\eta$ should be on order of
the mean free path $\ell$, so that $\eta\sim\kappa$. We have verified
this relation by plotting $\eta$ measured directly from thermal
profiles, as a function of the extrapolated temperature $T$ at the
boundary. These are summarized in Fig. \ref{fig18} (top). In the
figure we display only the jumps on the low temperature end since the
errors are much smaller. (The large gradient data on the high
temperature end, while consistent with the low end data, is quite
noisy.) We also show a fit to the data (dashes), which gives
\begin{equation}
\eta(T)= (6.1\pm0.5) T^{-1.5\pm0.1}.
\end{equation}
On the same figure we also show the behavior of $\kappa(T)$, which
indicates that $\eta\sim \kappa$ is consistent with the observed
behavior.

An independent verification of the behavior of the jumps can be
made by studying how the jumps depend on the heat flux. We let
$\eta=\alpha\kappa$, where $\alpha$ is a constant to be
determined.  {}From Eq. (\ref{eq:b-jump}), we can then associate
the heat flux with the right side of Eq. (\ref{eq:b-jump}):
\begin{equation}\label{t-jumps}
  T_i-T_i^0 \simeq \alpha \langle {\cal T}^{01}\rangle_{NE} 
          \sim    \alpha (T_2^0-T_1^0)\frac{\kappa(T_{av})}{L} 
          + \cdots.
\end{equation}
By plotting the boundary jumps directly as a function of the heat flux
in non-equilibrium steady states both near and far from equilibrium,
we obtain the data in Fig. \ref{fig18} (bottom). One can see that
there is a simple relationship, where the slope gives $\alpha=2.6(1)$,
consistent with the assumption that $\eta=\alpha\kappa$.  The
understanding of these jumps together with that of the temperature
profile~(\ref{t-profile}) provides a complete description of
$T(x)$ in terms of the boundary temperatures, $(T_1^0,T_2^0)$.
As was alluded to earlier, these simple
relations~(\ref{eq:b-jump}),(\ref{t-jumps}) break down in the LNE
region of Table 1, where the thermal gradients are too strong.

A note on boundary conditions is in order: There is reasonable
freedom in how the thermostats are implemented.  We can vary
the number of thermostatted sites or how the demons are couple to the
physical degrees of freedom.  We find that the changes in the way
thermostats are implemented bring about changes in the boundary jumps
when we are sufficiently far from equilibrium.  This will also result
in a different $\langle {\cal T}^{01}\rangle$ in a manner consistent
with Eq.~(\ref{ne-j}).  Different thermostats will correspond to
different values of $\alpha$, which is not an intrinsic parameter of
the $\phi^4$ theory but rather reflects different manners in which a
heat bath might efficiently couple to the scalar field theory.  In
other words, as expected, different thermostats do not change our
understanding of the physics at all.

In understanding the generality of Eq. (\ref{t-jumps}), we note that
the FPU $\beta$ model~(\ref{fpua}) displays temperature profiles that
depend on $L$. In Refs. \cite{lepri}, it has been shown that the heat
flux varies with the size as $ \left\langle{\cal
    T}^{01}\right\rangle\sim L^{-0.55}$.  Consequently, one would
expect that the boundary jumps seen there behave as $\delta T\sim
L^{-0.55}$.  A cursory analysis of those thermal profiles suggests
that this is indeed the case.
\subsection{ Non-Equilibrium Distributions}
It is worth making a few remarks on the non-equilibrium distribution
functions we have seen. 
Many usual approaches to the non-equilibrium statistical properties of 
systems make from the outset certain model dependent choices. In some
thermo-field approaches, the temperature profile $T(x)$ is chosen
to have a specific form, resulting in a statistical operator of the type
$\exp (-\tilde H(\pi ,\phi )/T(x))$, where $\tilde H(\pi,\phi)$ might
include contributions due to transport\cite{thermo-boson}. Employing the
approach of Zubarev\cite{zubarev}, one assumes a particular behavior
for the steady state distribution, which is then used to solve the
dynamical equations.

\begin{figure}
\begin{center}
    \leavevmode
  \epsfxsize=12cm\epsfbox{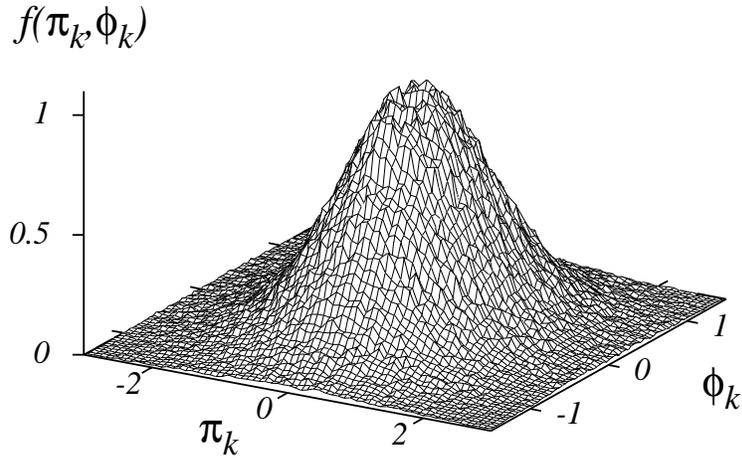}
  \caption{(Top) Non-equilibrium steady-state distribution
    function $f(\pi_k,\phi_k)$ for a system with
    $(T_1^0,T_2^0)=(0.1,6)$, where $k(=L/2)$ is taken to be the middle
    of the system. }
\label{fig17}
\end{center}
\end{figure}

\begin{figure}
\begin{center}
    \leavevmode
  \epsfxsize=12cm\epsfbox{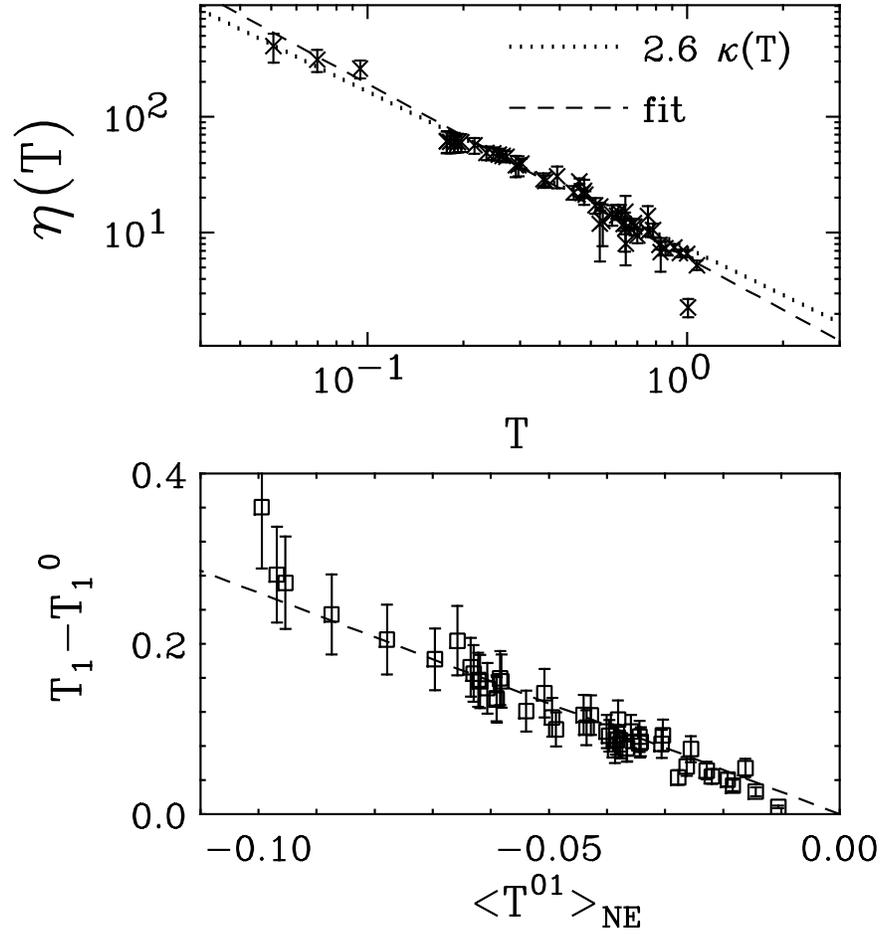}
  \caption{(Top) Behavior of boundary jumps as a function of
    temperature. $\eta(T)$ can be seen to obey a power law
    behavior (dashes) similar to the conductivity. (Bottom)
    Evidence that the boundary jumps are  simply related to the
    heat flux.}
\label{fig18}
\end{center}
\end{figure}

The local non-equilibrium distribution, $f^{(1)}_B(\phi_k,\pi_k)$ is
shown in Fig. \ref{fig17} for a system very far from equilibrium.
While the distribution $f$ in the full configuration space is expected
to be fractal, 
$f^{(1)}_B(\phi_k,\pi_k)$ is only a projection on a lower dimensional
space, and should be smooth.  Such a non-equilibrium statistical
operator will differ from those assumed in non-equilibrium approaches
in their correlations. One aspect of the fractal nature of $f$
is that the dynamical space is of reduced dimension, so
that additional correlations will exist that are not present in those
standard non-equilibrium approaches.

Another limit in which one obtains unusual thermal profiles is a
`ballistic' limit in which the mean free path is on the order or
larger than the size of the system. Because the excitations pass
through the system rather readily, one has large boundary jumps
on both ends. As a consequence, the thermal profile is almost
flat, in spite of the values of the endpoint temperatures. In
this case the system is still thermalized; the distributions of
momenta at the endpoints and inside are  gaussian. These
types of results are included in our figures of entropy,
specific heat and so forth, in the near-equilibrium data set,
and they fall in line with non-ballistic results.
\subsection{Local Equilibrium}
While we do not yet have any precise criteria regarding when local
equilibrium fails to be a good approximation, we can analyze various
physical quantities and compare them to their values when local
equilibrium holds.  This should provide us with a measure of the
dependence of the physics on local equilibrium.  While field dependent
observables, such as moments of $\phi(x)$ will depend on the model
under investigation, the momenta should be more robust.  A natural
place to start is with the analysis of cumulants. If local equilibrium
is expected holds, we should have
\begin{equation}
 \langle \pi_k^2\rangle = T,\qquad   \langle \pi_k^4\rangle = 3
 T^2, \qquad  \langle \pi_k^6\rangle = 15T^3
\end{equation}
and so forth. Equivalently, we can say that
\begin{equation} 
  \llangle \pi_k^4\rrangle =\langle \pi_k^4\rangle - 3
  \langle\pi_k^2\rangle^2 ,\qquad
  \llangle \pi_k^6\rrangle =\langle \pi_k^6\rangle - 15
  \langle\pi_k^4\rangle   \langle\pi_k^2\rangle 
  +30\langle\pi_k^2\rangle ^3.
\end{equation}
vanish in local equilibrium. Hence one measure of the breaking of
local equilibrium when we are far from equilibrium is
\begin{equation}
  \label{eq:cumulant}
   \frac{\llangle \pi_k^4\rrangle}{3  \langle \pi_k^2\rangle^2} = 
       \frac{ \langle \pi_k^4\rangle}{ 3 \langle\pi_k^2\rangle^2} -1.
\end{equation}
Similar expressions for higher cumulants are also possible. 

Non-equilibrium behavior begins to emerge in certain quantities as
soon as $T_1\not=T_2$. For instance, $\langle\pi^4(x)\rangle/\langle
\pi^2(x)\rangle^2\geq 3$, the equality holding in equilibrium, and the
value growing as one departs from equilibrium. One can also see that
the non-equilibrium measure is not locally Boltzmann since quantities
such as $\langle\pi(x)\phi(x')\rangle\not=0$ for $x\not=x'$, as is
expected for a system supporting some type of transport. However, the
origins in this case we attribute not to additional terms in the
statistical distribution $f$, but rather to additional correlations in
non-equilibrium measure due to its reduced dimensionality, as expected
from dynamical systems theory.
Unfortunately, there is no way to explicitly estimate the dimensional
loss without direct measurement of the entire Lyapunov spectrum, which
is a rather numerically intensive task.

\begin{figure}
\begin{center}
    \leavevmode
    \epsfxsize=12cm\epsfbox{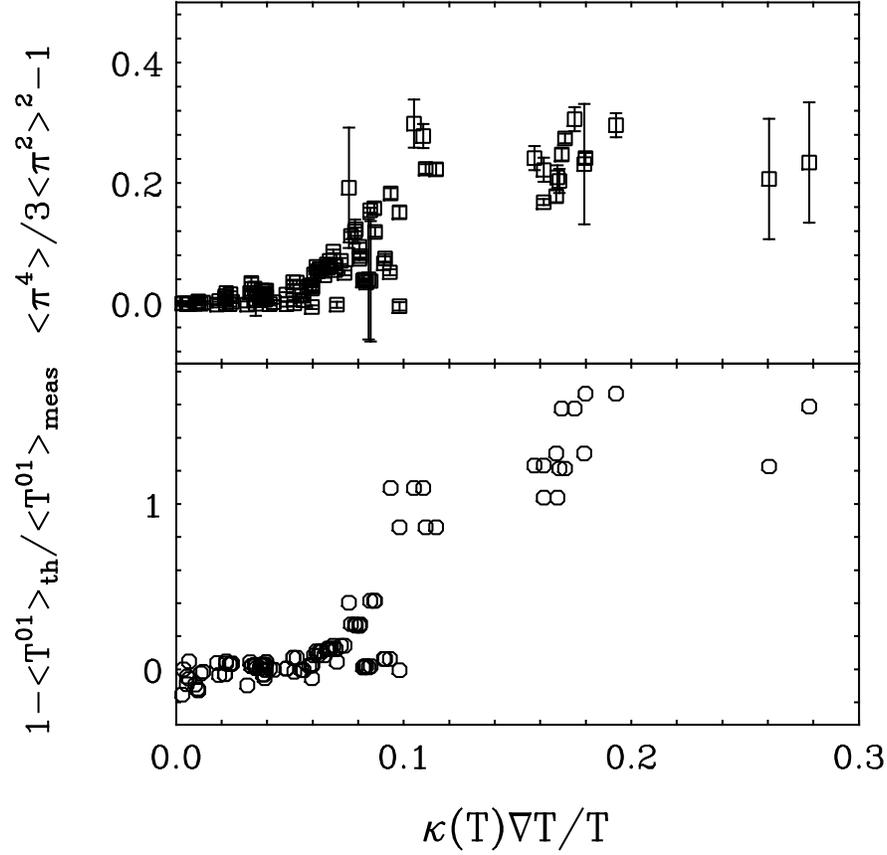}
    \caption{Measures of the breaking of local equilibrium. (Top) The
      behavior of the cumulants of the momentum distribution for
      various non-equilibrium steady states, as a function of $\kappa
      \nabla T/T\sim \ell\nabla T/T$, where $\ell$ is the mean free
      path. One can see that noticeable departures from gaussian
      momentum distributions develop for $\kappa \nabla T/T\gsim 1/10$.
      (Bottom) Departure of the predicted heat flux from the measured
      heat flux. One can see that the predictions, based on local
      equilibrium and the shape of the thermal profile, begin to
      deviate from measured values in the same regime.}
  \label{fig:lne}
\end{center}
\end{figure}

As one moves away from equilibrium, the linear response theory is
expected to eventually breakdown.  Such behavior is confirmed for our
system; when the temperature gradient becomes too large, the formula
for the heat flow (\ref{ne-j}) ceases to be valid.  We display the
relative difference of the measured current to the current obtained
from the linear response theory (\ref{ne-j}) in
Fig.~\ref{fig:lne}~(bottom).  It can be seen that the relative deviations
can be of order one for large thermal gradients signaling the
breakdown of the linear response theory.  The deviations are plotted
against $\kappa(T)\nabla T/T$ which seems to be the natural scale,
since $\kappa$ is the mean free path, roughly speaking, which is the
natural length scale in the problem as discussed in
section~\ref{sec:non-eq}~D.  One obvious possibility would be to
interpret this as a nonlinear response of the thermal conductivity,
$\kappa(T, {\cal T}^{01})$, which may, for instance, be parametrized
as
\begin{equation}
  \label{nonlinear-response}
  \kappa(T,{\cal T}^{01}) = 
  \kappa_0(T) + \kappa_2(T) \left({\cal T}^{01}\right)^2 
  + {\cal O}\left(\left({\cal T}^{01}\right)^4\right) .
\end{equation}
Such approaches have been discussed in the literature \cite{revs}.

{\it However,} such an interpretation presupposes, perhaps tacitly,
that the concept of local equilibrium holds and that the standard
notion of temperature applies, amongst other things.  Therefore, it is
imperative to first check that the local equilibrium is achieved in
this `non-linear' regime and that it is this point that we now wish
to investigate with care.  Such questions have been asked previously
and in those situations, the local equilibrium was seen to be valid
\cite{local-eq}.
First, we need to ask ourselves what constitutes local
equilibrium?  The concept of local equilibrium has been defined
previously \cite{zubarev,mclennan}\ and it is not our intention
here to discern the possible differences in the various
definitions of local equilibrium.  In the least, local
equilibrium assumes that we have a Maxwellian distribution for
the momentum for the class of Hamiltonians we work with, leading
to the usual concept of temperature, which will be our point of
investigation.

\begin{figure}
\begin{center}
    \leavevmode
  \epsfxsize=12cm\epsfbox{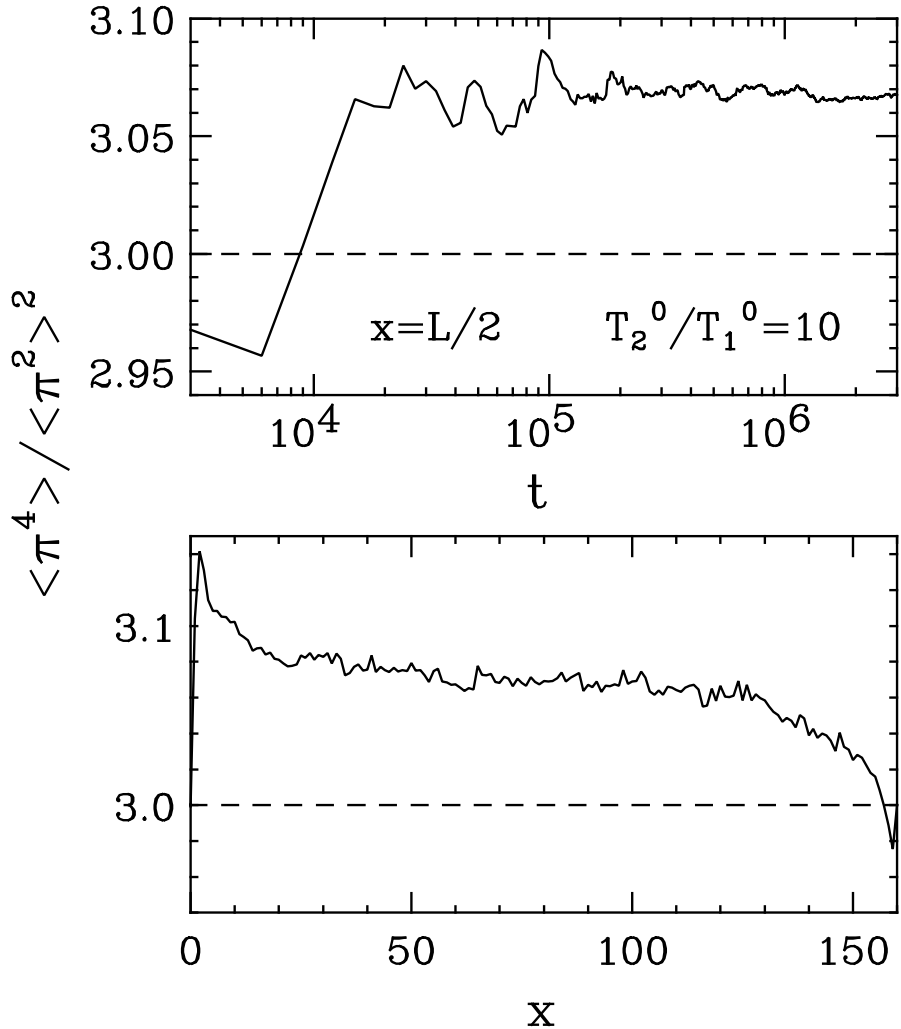}
  \caption{Behavior of the fourth moment $\langle \pi^4\rangle$
    normalized by $\langle\pi^2\rangle^2$. (Top) We show the time
    evolution at the center of a system with $L=160$ sites and
    boundary temperatures $(T_1^0,T_2^0)=(0.1,1)$. We see that the
    ratio is well converged. (Bottom) We show the behavior of the
    moments along the lattice. The local equilibrium value is
    indicated by the dashed line. While the endpoints are in thermal
    equilibrium one can see that the low temperature end suffers the
    largest departures from local equilibrium. }
  \label{fig:cumulants}
\end{center}
\end{figure}

In Fig.~\ref{fig:lne}~(top), we plot the fourth cumulant of $\pi$,
(\ref{eq:cumulant}), against $\kappa(T)\nabla T/T$.  The cumulant,
which is defined {\it locally}, quite clearly deviates from the
equilibrium value under strong thermal gradients.  Even in these
situations, the thermostatted boundary sites {\it are} in local
equilibrium, as they should be, as shown in Fig.~\ref{fig:cumulants}.
It can be seen that the deviations from local equilibrium start to
occur at roughly the same value of $\kappa(T)\nabla T/T\sim 1/10$
where linear response theory breaks down.   
In Table 1, we  identify this as 
a steady state which is locally non-equilibrium (LNE). 
We have verified that higher cumulants display similar behavior. 
Therefore, at least in our model, the breakdown of local equilibrium
needs to be considered when nonlinearity of the response is to be
analyzed.  Of course, this does not preclude the possibility of the
nonlinearity of the response as discussed above, but that the
nonlinearity needs to be disentangled from the deviations from local
equilibrium with care.

We emphasize here that {\it \`a priori}, this needs not be the case.
Namely, it is in principle possible that there is a region where the
concept of local equilibrium is still valid and yet the linear
response theory breaks down.  In such a situation, `non--linear
response' theory should be quite appropriate for analyzing the
situation.  However, in the cases we studied, such regimes do not
exist and the failure of the linear response theory occurs
simultaneously with the breakdown of local equilibrium.
\section{Conclusions}
We have constructed non-equilibrium steady states for classical
$\phi^4$ lattice field theory in one dimension, under conditions near
and far from equilibrium.  We obtained the behavior of the thermal
conductivity with respect to the temperature in the linear regime and
found that the results were consistent with the linear response
regime.  The underlying dynamics of the theory was investigated and
physical quantities such as the speed of sound, heat capacity,
Boltzmann's entropy and their temperature dependence, were obtained.
The results could consistently understood using the kinetic theory
approach.  This understanding was further used to clarify the dynamics
behind the temperature gaps that arise at the boundaries of the
system.  We also found that for temperature gradients that are not too
large, the linear response law is adequate for understanding the
behavior of the system, even though the temperature profile might be
visibly non-linear.  For even larger gradients, even though the system
is in a steady state, we found that the linear response law eventually
ceases to hold, but the local equilibrium is also violated.

We have classified the steady states in Table 1, which identify
distinct dynamical regimes of the theory. It would be nice to develop
more precise measures for these dynamical regimes, but it is clear
that even a one component classical, lattice field theory does contain
a means to understand the non-equilibrium physics of many-body
systems. It would be interesting to extend these results to theory
with phase transitions, multi-component theories which would also
allow the analysis of Onsager reciprocity relations and so on. The
additional degrees of freedom will provide additional measurable
quantities, but we expect many of the qualitative features of this
simple model to persist.

We acknowledge support through 
the grants from Keio University and DOE grant DE-FG02-91ER40608.  We
would like to thank Guy Moore and Larry Yaffe for enlightening
discussions and the the Institute for Nuclear Physics at University of
Washington for hospitality, where some of the work was conducted.


\end{document}